\documentclass[sigconf]{acmart}
\usepackage{multirow}
\usepackage{booktabs}
\usepackage[normalem]{ulem}
\usepackage{algorithm}
\usepackage{algorithmic}

\AtBeginDocument{%
  \providecommand\BibTeX{{%
    \normalfont B\kern-0.5em{\scshape i\kern-0.25em b}\kern-0.8em\TeX}}}

\copyrightyear{2023} 
\acmYear{2023} 
\setcopyright{acmlicensed}\acmConference[KDD '23]{Proceedings of the 29th ACM SIGKDD Conference on Knowledge Discovery and Data Mining}{August 6--10, 2023}{Long Beach, CA, USA}
\acmBooktitle{Proceedings of the 29th ACM SIGKDD Conference on Knowledge Discovery and Data Mining (KDD '23), August 6--10, 2023, Long Beach, CA, USA}
\acmPrice{15.00}
\acmDOI{10.1145/3580305.3599411}
\acmISBN{979-8-4007-0103-0/23/08}

\begin{document}

\title{Learning to Relate to Previous Turns in Conversational Search}

\author{Fengran Mo}
\orcid{0000-0002-0838-6994}
\author{Jian-Yun Nie}
\orcid{0000-0003-1556-3335}
\affiliation{%
  \institution{University of Montreal}
  \city{Montreal}
  \state{Quebec}
  \country{Canada}
}
\email{fengran.mo@umontreal.ca}
\email{nie@iro.umontreal.ca}

\author{Kaiyu Huang}
\orcid{0000-0001-6779-1810}
\affiliation{%
  \institution{Institute for AI Industry Research, Tsinghua University \city{Beijing}
  \country{China}}}
\email{huangkaiyu@air.tsinghua.edu.cn}

\author{Kelong Mao}
\orcid{0000-0002-5648-568X}
\affiliation{%
  \institution{Gaoling School of Artificial Intelligence, Renmin University of China \city{Beijing}
  \country{China}}
}
\email{mkl@ruc.edu.cn}

\author{Yutao Zhu}
\orcid{0000-0002-9432-3251}
\affiliation{%
 \institution{University of Montreal}
 \city{Montreal}
 \state{Quebec}
 \country{Canada}}
\email{yutaozhu94@gmail.com}

\author{Peng Li}
\orcid{0000-0003-1374-5979}
\affiliation{%
  \institution{Institute for AI Industry Research, Tsinghua University \city{Beijing} \country{China}
  }}
\email{lipeng@air.tsinghua.edu.cn}

\author{Yang Liu}
\orcid{0000-0002-3087-242X}
\affiliation{%
  \institution{Department of Computer Science, Tsinghua University \city{Beijing} \country{China}
  }}
\email{liuyang2011@tsinghua.edu.cn}

\renewcommand{\shortauthors}{Mo et al.}

\begin{abstract}
Conversational search allows a user to interact with a search system in multiple turns. A query is strongly dependent on the conversation context. An effective way to improve retrieval effectiveness is to expand the current query with historical queries. However, not all the previous queries are related to, and useful for expanding the current query. In this paper, we propose a new method to select relevant historical queries that are useful for the current query. To cope with the lack of labeled training data, we use a pseudo-labeling approach to annotate useful historical queries based on their impact on the retrieval results. The pseudo-labeled data are used to train a selection model. We further propose a multi-task learning framework to jointly train the selector and the retriever during fine-tuning, allowing us to mitigate the possible inconsistency between the pseudo labels and the changed retriever. Extensive experiments on four conversational search datasets demonstrate the effectiveness and broad applicability of our method compared with several strong baselines. 
\end{abstract}




\maketitle

\section{Introduction}

Conversational search~\cite{gao2022neural} is an important emerging branch of information retrieval following the rapid development of intelligent assistants  (e.g., Siri and Alexa). Compared with the traditional keyword-based ad-hoc search, conversational search allows the user to interact with the system in multiple turns to seek information via a conversation in natural language. Not only this is a natural way for users to interact with an IR system, but also it has the potential to deal with complex information needs~\cite{dalton2022conversational}. 

As in any conversation in natural language, queries in conversational search may involve omissions, references to previous turns, and ambiguities~\cite{radlinski2017theoretical}. Thus, a primary challenge for effective conversation search is to determine the underlying information need by understanding context-dependent query turns.
Some previous studies~\cite{voskarides2020query,vakulenko2021question,yu2020few,2020Making,wu2021conqrr} exploited a two-stage pipeline, which first trains a query rewriting (QR) model with external data or human reformulated queries, then feeds the rewritten queries into a retriever. However, it is required that sufficient manual annotations are available for model training, which is difficult to obtain in practice~\cite{anantha2021open,adlakha2022topiocqa}. Furthermore, the manually reformulated queries may not be the best search queries because the reformulation is based on human understanding of the query rather than on the search results. We also observe that the rewriting model is usually trained separately from the retriever, making it difficult to optimize the whole system for the final retrieval performance~\cite{lin2021contextualized,wu2021conqrr}. In particular, the learned rewriting model may not best fit the  retriever used in the subsequent step.

\begin{figure}[t]
\centering
\includegraphics[scale=.45]{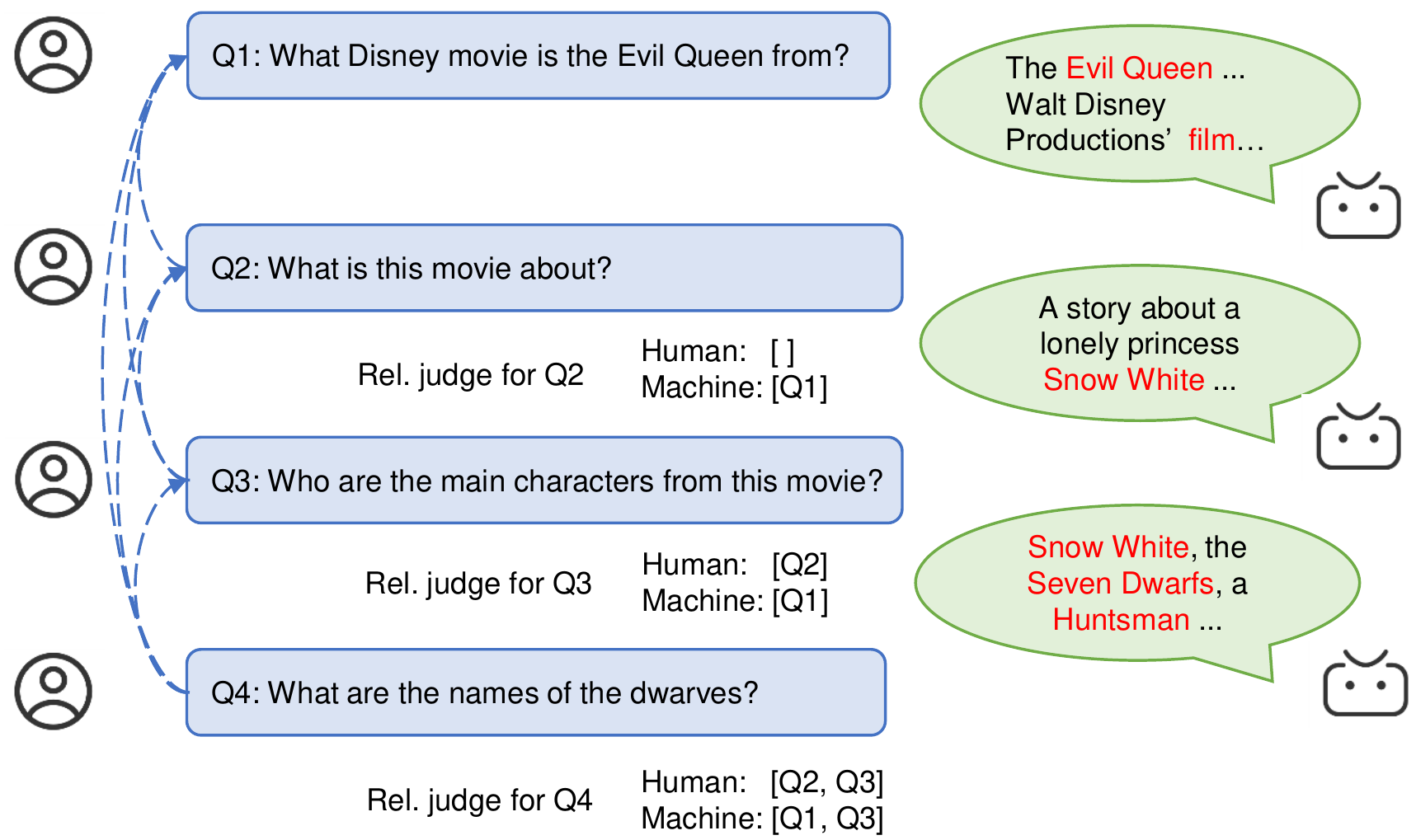}
\caption{$Q2$, $Q3$ and $Q4$ are about "Snow White", while  $Q1$ is about "Evil Queen". $Q2$ and $Q3$ are judged relevant to $Q4$ in human judgments. Instead, our model judges $Q1$ and $Q3$ are useful for $Q4$ based on their impact to improve the retrieval effectiveness of $Q4$.}
\label{fig: example}
\vspace{-4ex}
\end{figure}

Another research direction~\cite{qu2020open,yu2021few,kim2022saving,mao2022curriculum} tries to train a conversational query encoder by leveraging all the previous context information.
A common approach uses all the historical queries in the session to expand the current query. 
Although this expansion leads to improved results, it makes a strong assumption that all the previous queries are related to the current one. In reality, we often observe topic switches within a conversation session, which means the current query may not be related to the previous ones.
Fig.~\ref{fig: example} illustrates such an example, where we see that only part of the historical queries is  relevant to the last query (either by human annotators or by our model). Simply leveraging all the historical queries will inevitably inject irrelevant information into the expanded query and result in sub-optimal queries. This fact will be experimentally confirmed in Sec.~\ref{sec: Re-examination}. 
Intuitively, a better solution that we will implement is to use only the relevant previous turns to expand/rewrite the current query.
So the crucial question is: \textit{how can we select useful information from history context for query expansion/reformulation?}

A naive idea would be training a classification model based on human annotations to identify the relevant previous queries. Unfortunately, this approach cannot be implemented due to the fact that human annotations of query relevance are usually unavailable in practice.
Such relevance labels are also unavailable in most existing datasets for conversational search~\cite{dalton2020trec,dalton2021cast,anantha2021open,adlakha2022topiocqa}. Thus, identifying relevant previous queries is a non-trivial task in the absence of relevance labels.

Some existing studies~\cite{qu2020open,kim2022saving,li2022dynamic,yu2021few,lin2021contextualized} exploit the powerful pre-trained language models (PLM)~\cite{devlin2019bert,liu2019roberta} and use the attention mechanism~\cite{vaswani2017attention} to determine how strongly historical queries are related to the current one. A common practice is to fine-tune a pre-trained dense retriever~\cite{xiong2020approximate,karpukhin2020dense} with conversational search data. This approach would require a considerable amount of annotated conversational search sessions, with relevance labels for previous queries. 
To this end, some pioneering studies~\cite{adlakha2022topiocqa,mao2022curriculum} proposed annotated conversational search sessions.
However, this gives rise to two critical issues: (1) It is difficult to produce a large number of manual annotations in general conversation search sessions; and (2) it is questionable if human annotations are the best supervision signal to use to train a model. 
In fact, when a human annotates the relevance of a previous query, she/he would basically consider if the two queries are semantically related, without being able to judge if expanding the query with it would improve the search result. This latter would require deep knowledge about how the retrieval model will use the query, which often remains opaque to a human annotator. Therefore, a human annotator could select seemingly relevant but useless queries, while missing some semantically less related but actually useful queries.

In this paper, we investigate how to select \textit{useful} historical turns to expand the current query in conversational  search. The selection is made according to the usefulness of the previous queries, i.e. if the latter can help improve the retrieval effectiveness of the current query.
Intuitively, this could help select more useful previous queries for the search task. As we can see in Fig.~\ref{fig: example}, our system can make different judgments than a human annotator. Q1 is judged useful because it is about the same movie, although about a different character. This is a piece of useful information to expand Q4. On the other hand, Q2, which is manually judged relevant, does not provide any useful information although being about the same character (``Snow White'').

To select useful previous turns, we are inspired by earlier work that selects useful expansion terms or pseudo-feedback documents for query expansion according to their impact~\cite{cao2008selecting,he2009cikm}. Our situation is similar: we consider the previous turns as candidates for expanding the current query and we aim at selecting the ones that can improve the search effectiveness. 
Therefore, a similar principle can be used, i.e., a selection model is trained with a set of automatic labels generated according to whether the previous turn can bring improvement in retrieval. 

We also observe that most previous approaches separately train a query expansion/rewriting model and a retrieval model. In fact, these two components are strongly dependent - the change in one would impact the other. To jointly optimize both, we design a multi-task learning method. 
We conduct extensive experiments on four widely used conversational search datasets.
Experimental results demonstrate the effectiveness of our proposed method, showing significant improvements over various baselines. 

Our contributions are summarized as follows:

(1) We propose an efficient and effective method to select useful previous queries to expand the current query without human annotations.

(2) We propose a multi-task learning method to jointly optimize the query selection model and the dense retrieval model.

(3) We demonstrate the effectiveness of the proposed method and show its broad applicability in various settings (i.e. with or without training data). In addition, the proposed method can select better previous queries than human annotators.

\section{Related Work}
\textbf{Conversational Search.}
Conversational search is defined as iteratively retrieving documents for user's queries in a multi-round dialog~\cite{dalton2020trec,dalton2021cast}. A later query may depend on the previous (historical) ones in the session. Thus the major challenge is to formulate a de-contextualized query for a standard search system according to the historical context. 

One research direction is to explicitly reformulate queries. 
Typical methods include selecting important terms from the previous queries to expand the current query via a term classifier~\cite{2020Making,voskarides2020query}, or rewriting the query based on a generative pre-trained language model (PLM)~\cite{yu2020few,wu2021conqrr,mao2023large,mo2023convgqr}. Both methods need external data or human-rewritten oracle queries for training.
However, it is found that the manual queries may be sub-optimal for search~\cite{wu2021conqrr,lin2021contextualized}. Therefore, training a query reformulation/rewriting model according to manually reformulated queries may lead to a sub-optimal solution.

Another approach is to fine-tune a dense retriever using conversational search data~\cite{qu2020open,kim2022saving,li2022dynamic,mao2022convtrans,mao2023learning}. Since there are limited supervision signals in the current conversational search datasets, the dense retriever will have difficulty to identify the relevant  historical turns to enhance the current query.  
To address this problem, some few/zero-shot learning methods have been proposed~\cite{yu2021few,krasakis2022zero,mao2022curriculum}. They try to adapt the ad-hoc search retriever to the conversational scenario with external resources.
Nevertheless, these approaches do not contain an explicit selection of relevant historical turns for the current query. As we stated earlier, this selection is critical for conversational search.

\textbf{Query Expansion.}
For conversation search, expanding the query in the current turn based on historical context can be viewed as a type of query expansion~\cite{efthimiadis1996query}, which is an effective technique for improving retrieval performance.
The expansion terms can come from different resources such as user logs, pseudo-relevant feedback, and document collection. It is critical to use appropriate terms to expand the query. However, it has been shown that the candidate expansion terms selected by a method (e.g. pseudo-relevance feedback) are very noisy: they could be useful, useless, or even harmful to improve the retrieval effectiveness.
Therefore, it is important to select only good terms for query expansion~\cite{cao2008selecting,sordoni2015hierarchical}. In the case of pseudo-relevance feedback, there is no further indication or annotation about the usefulness of an expansion term. Therefore, \citet{cao2008selecting} proposed a method to automatically annotate the usefulness of a candidate term by testing if incorporating it into an expanded query. 
If it does, then the term is labeled useful; if it degrades the performance, then it is labeled harmful. A similar approach has been adapted for selecting pseudo feedback documents for query expansion \cite{he2009cikm}. 
Besides, using entity overlap to select the history queries is another interesting naive solution. However, previous works~\cite{vakulenko2018measuring,joko2021conversational} found that entities are often omitted in later turns rather than repeated, which results in poor coverage in finding useful turns in conversational search.
In our work, we borrow the idea of pseudo-labeling to annotate the usefulness (or relevance) of a historical query. Based on the pseudo labels, a query selection model (selector) will be trained.

\section{Task Formulation}
\label{sec: Task Formulation}
In conversational search, we are given a multi-turn conversation corresponding to a sequence of queries $Q = \{q_i\}_{i=1}^n$. The goal is to find the relevant passages $p^+$ from a collection $D$ for the last query $q_n$.
Each query turn can be ambiguous and is usually context-dependent, thus we need to understand its real search intent based on its conversational context $C$. Several types of context $C$ have been considered in previous works: (1) The context contains historical queries and the corresponding ground-truth passages (answers) $C = \{q_i,p_i^*(a_i^*)\}_{i=1}^{n-1}$~\cite{vakulenko2021question,krasakis2022zero,kim2022saving,mao2023learning} and (2) it contains historical queries only $C = \{q_i\}_{i=1}^{n-1}$~\cite{qu2020open,yu2021few,mao2022curriculum}. Despite the fact that the first type of context can lead to better results, it is believed~\cite{mandya2020not,siblini2021towards,li2022ditch} that it is unrealistic to assume all historical queries are answered correctly.
In a realistic scenario, the answers are provided by the system and are not always correct. Therefore, we consider the second type of context in this paper, i.e. our context contains only the historical queries $C = \{q_i\}_{i=1}^{n-1}$. Notice, however, that the method we propose can be easily extended to the first type of context.

\section{Re-examination of Conversational Query  Expansion}
\label{sec: Re-examination}
The general assumption behind the existing conversational dense retrieval methods~\cite{vakulenko2021question,li2022dynamic,krasakis2022zero,qu2020open,yu2021few,lin2021contextualized} is that \textit{all historical queries in previous turns are useful for understanding current query.} Therefore, expanding the query by all of them should yield better retrieval results.
Typically, they reformulate the current query by concatenating all historical queries.
Despite the fact that such an expanded query leads to improved performance (compared to no expansion), we argue that the strategy is sub-optimal. On one hand, the historical queries may be on a different topic, thus not useful to enhance the current query. This happens especially in long sessions where the user may be interested in different topics. On the other hand, the historical queries are not of the same usefulness for expanding the current query. Some may be more related than others. 

The above phenomena have been experimentally shown in \cite{cao2008selecting} when expanding a query in pseudo-relevance feedback. To verify if the same phenomena exist in conversational search, we run a similar experiment to compare the following two scenarios: (1) expand the current query with all historical queries, (2) expand it with only relevant historical queries/terms. We want to demonstrate that the second strategy is much better than the first one. 
Notice that the first method is widely used in existing works. The second one is an idealized case because the relevance labels of previous queries are usually unavailable.
Meanwhile, within the second strategy, we want to compare the effectiveness between token-level and turn-level expansion based on different types of retrievers.
Following \cite{cao2008selecting}, we identify the useful expansion queries from the history according to whether the selected expansions can bring improvement in retrieval. We describe the process below.

\subsection{Pseudo Relevance Labeling}
\label{sec: PRL}

Our goal is to generate gold Pseudo Relevance Labels (PRL) for each historical query/term with respect to a given query. We assume that a relevant historical turn/term should make a positive impact when the current query is expanded with it. The pseudo labeling is based on the impact on retrieval results of a candidate turn/term when the latter is used to expand the query. 

The turn-level generation procedure is introduced in Algorithm~\ref{alg: PRL}. We first use the current raw query to retrieve the ranked list of documents and calculate the base retrieval score (MRR in this case). Then we iteratively concatenate the query with each of the historical queries as a new reformulated query and get the corresponding retrieval score. If the second score is higher than the initial one, then the relevance of the historical query is labeled ``positive'', otherwise, ``negative''. Query expansion is produced by concatenating the expansion query $h_i$ with the current one $q_n$, i.e. $q^{exp} = q_n \circ h_i$, where $\circ$ means concatenation. 
In the same way, we can also create term-level pseudo labels for all the terms in the historical queries.

\begin{algorithm}[t]
	\caption{Generating Gold Pseudo Relevant Label (PRL)}
	\renewcommand{\algorithmicrequire}{\textbf{Input:}}
	\renewcommand{\algorithmicensure}{\textbf{Output:}}
	\begin{algorithmic}[1]
		\REQUIRE The $n^{th}$ query turn $q_n$, historical queries $h_{1:n-1} (n>1)$, existing retriever $f$
		\ENSURE The $n^{th}$ turn gold pseudo relevant labels $L_n^{gold}$
		\STATE RankList $R_q$ = $f(q_n)$
		\STATE Evaluate $R_q \rightarrow$ base retrieval score $S_q$
		\STATE Initial list of pseudo relevance labels $L_n^{gold} = []$
		\FOR {$i$ from $1$ to $n-1$}
		    \STATE RankList $R_{h_i}$ = $f(q_n \circ h_i)$
		    \STATE Evaluate $R_{h_i} \rightarrow$ expansion retrieval score $S_{h_i}$
		    \IF {$S_{h_i} > S_{q_n}$}
		        \STATE Add positive label to $L_n^{gold}$
		    \ELSE
		        \STATE Add negative label to $L_n^{gold}$
            \ENDIF
        \ENDFOR
	\end{algorithmic} 
	\label{alg: PRL}
\end{algorithm}

\begin{figure}[t]
    \centering    
\includegraphics[width=\linewidth]{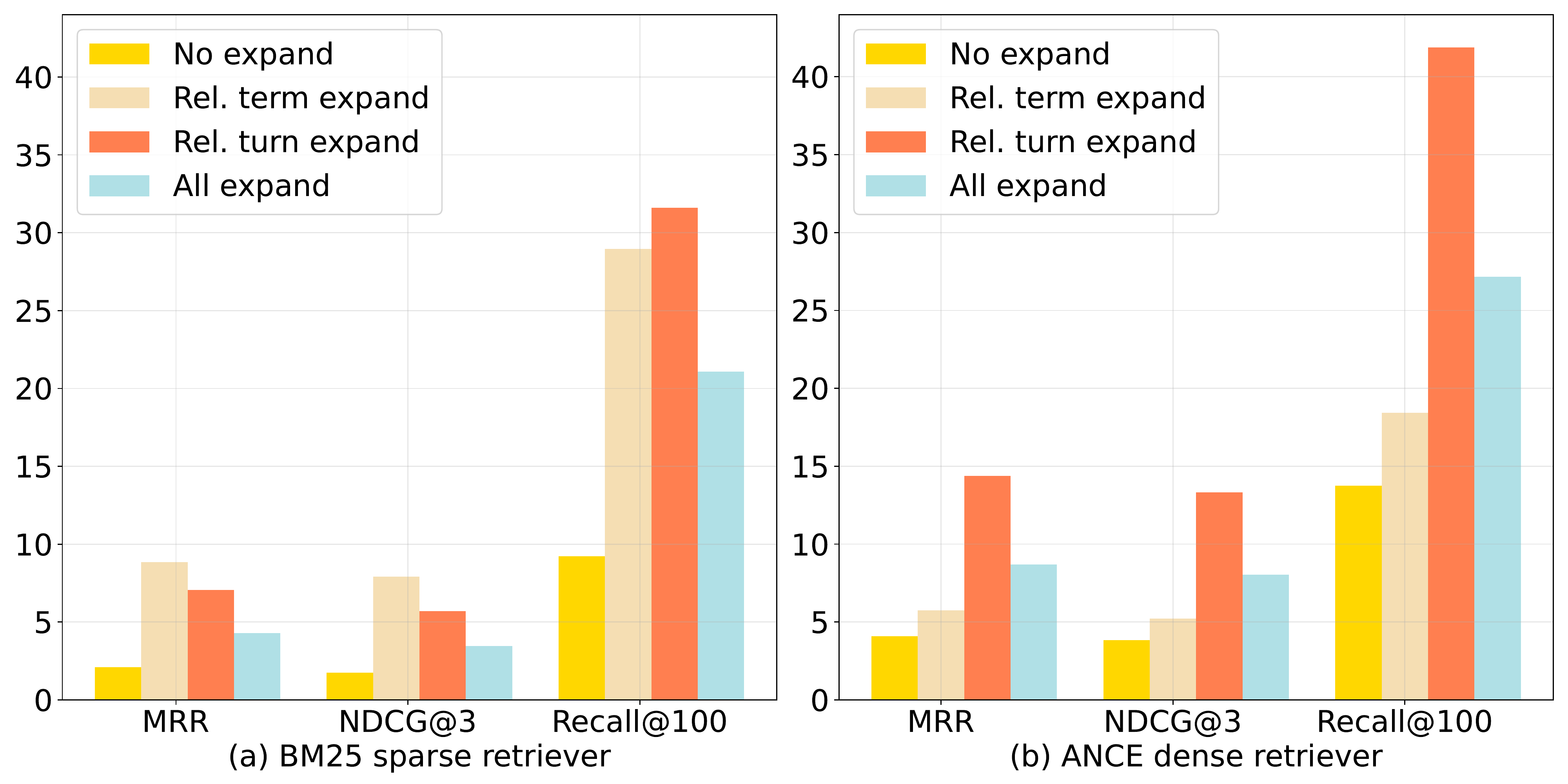}
    \caption{Potential effectiveness of expansion methods for retrieval with different experiment settings on TopiOCQA.}
    \label{fig: Effectiveness}
\vspace{-2ex}
\end{figure}

\subsection{Effectiveness of Selecting Useful Expansions}

We assume the gold PRL as the results from an oracle classifier that can correctly separate the useful and harmful historical expansion turns/terms. Before proposing an approach to select useful expansion turns/terms, we first examine if the gold PRL can indeed improve retrieval performance by using the relevant expansion turns/terms.
We formulate these query expansion forms 
as:
\begin{align*}
    q^{\text{raw}} &= q_n, \\
    q^{\text{all}} &= h_1 \circ \cdots h_{i} \cdots \circ h_{n-1} \circ q_n, \quad i \in [1, n-1], \\
    q^{\text{PRL}} &= h_1^L \circ \cdots h_{j}^L \cdots  \circ q_n, \quad  \quad  j \in [1, n-1] \ and \ L^{gold}_n[j] = 1.
\end{align*}

\begin{figure*}[t]
\centering
\includegraphics[width=0.85\textwidth]{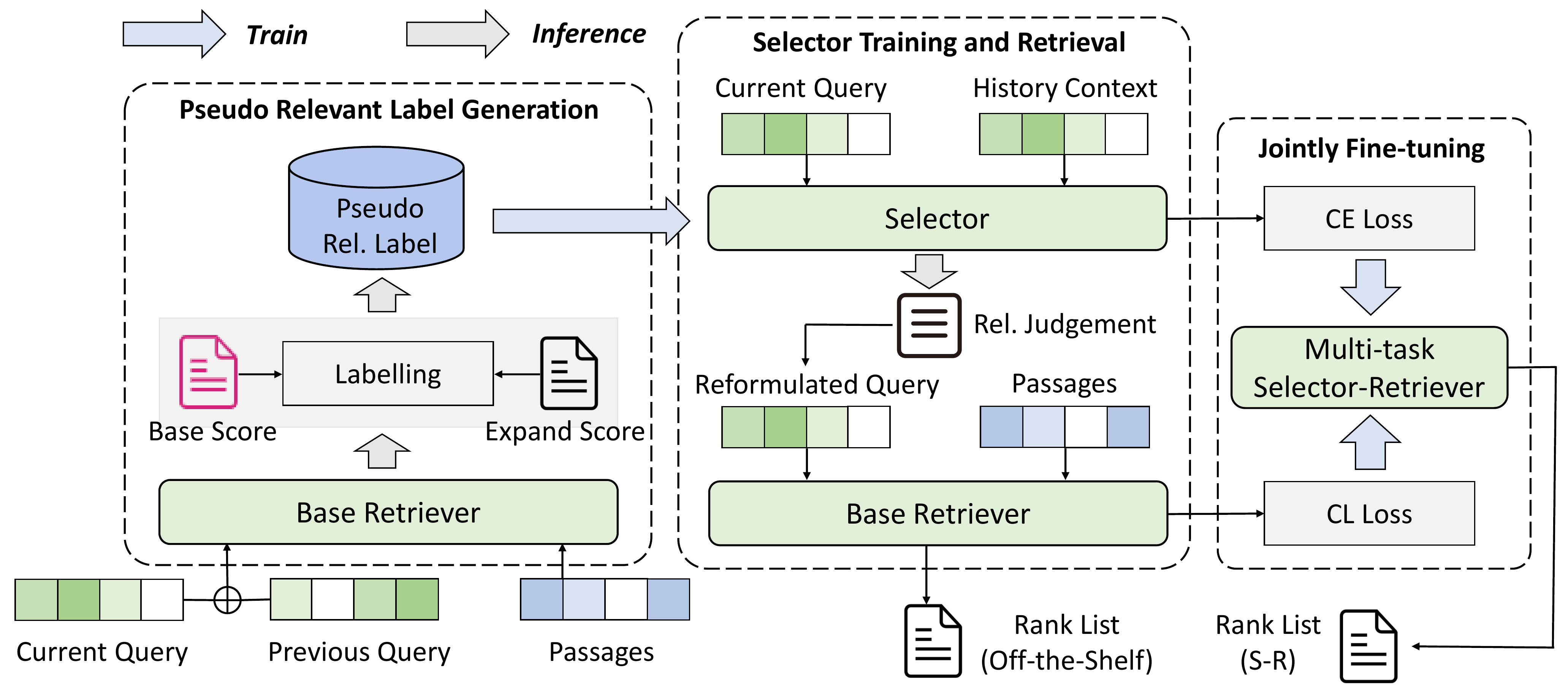}
\caption{Overview of our method and workflow. The PRLs are first generated through a base retriever (left part), then the PRLs are used to train the selector (middle part) or selector-retriever (S-R) jointly fine-tuning (right part). The relevant judgment produced by the trained selector can perform off-the-shelf retrieval and the fine-tuned (S-R) can act as a conversational retriever.}
\label{fig: Overview}
\vspace{-2ex}
\end{figure*}

Fig.~\ref{fig: Effectiveness} shows the retrieval effectiveness scores with these types of queries on the TopiOCQA dataset.
We can see significant improvement in retrieval effectiveness by expanding the queries with generated PRL with two retrievers: BM25 (sparse)~\cite{robertson2009probabilistic}, ANCE (dense)~\cite{xiong2020approximate}. 
In general, we see that expanding queries by all historical queries is beneficial, leading to improved effectiveness compared to the raw queries. This result is consistent with the observations in previous studies. However, when we use only relevant queries annotated by PRL, the effectiveness is much higher. This demonstrates clearly the potential benefit of selecting the relevant previous historical context and confirms our earlier assumption. 
We also find the expansion by the relevant tokens is more suitable  with the sparse retriever while the turn-level expansion performs better with the dense retriever. The comparison between the two retrievers clearly shows the advantage of the dense retriever ANCE.
Therefore, in our experiments, we will use ANCE as the retrieval backbone and relevant historical queries as expansion units.

Given the high impact of relevant queries for expansion, our problem is to develop an effective selection model to identify them. 
The principle we use is the selected expansion query can bring improvement in retrieval and be judged as relevant for the current query turn by a neural selection model.

\section{Methodology}

\subsection{Overview}
So far, the main approaches to QR~\cite{yu2020few,anantha2021open,wu2021conqrr,voskarides2020query} rely on external supervision signals, i.e. human-rewritten queries, to train a rewriting model. In practice, it is difficult to obtain such manually reformulated queries. Other methods expand the query with all historical queries~\cite{qu2020open,yu2021few,lin2021contextualized,kim2022saving}, which will inject noise during expansion, as we showed in the previous section.   
To overcome these issues, our method learns a query selection model (selector) based on the pseudo-labeled data to select useful historical queries for expansion and produces an expanded query to be submitted to a retriever.

However, we also notice that there is a hidden dependency between the retrieval model and the expansion process.
An off-the-shelf retriever such as ANCE has been fine-tuned with unexpanded queries. Fine-tuning it further for expanded queries can yield better results. Similarly, the selector also relies on a retriever to determine useful expansion queries. 
To account for the dependency, we design a multi-task learning framework to jointly fine-tune the selector-retriever (S-R) models through conversational search data. 

The overview of our methods is in Fig.~\ref{fig: Overview}, which contains three main components: PRLs generation, selector model training with off-the-shelf retrieval, and multi-task S-R model jointly fine-tuning.

\subsection{Selector: Selecting Useful Expansion Queries}

\subsubsection{Classification of Historical Queries}

The goal of the selector model is to identify if a historical query $h_i$ is relevant (useful) to the current query $q_n$. 
We use the pseudo-labeled queries as identified in Sec.~\ref{sec: Re-examination} as training data to train a selector model. 

\textbf{Selector Model.}
The purpose of the selector is to determine which historical queries are relevant (useful) for the current one. This can be done in two possible ways: as a classification problem or as a search problem. With a classification approach, we train a classifier to classify candidate queries into relevant/irrelevant classes. A straightforward approach is to directly train a powerful PLM such as Bert~\cite{devlin2019bert} with PRL data as a selector. 
The possible problem with this solution is that a Bert model without specific retrieval fine-tuning might not perform well on query encoding~\cite{qiao2019understanding}. Alternatively, we can also exploit a search tool to determine a matching score between the current query and a candidate. We leverage the ANCE~\cite{xiong2020approximate} dense retriever which has already fine-tuned by ad-hoc search data~\cite{nguyen2016ms} 
to train a new selector model. We will test both approaches in our experiments.

\textbf{Weight Assignment to Training Samples.}
The negative labels annotated automatically by Algorithm~\ref{alg: PRL} is ten times more than positive labels. 
However, in Sec.~\ref{sec: Re-examination} we showed that even expanding the query by all the historical queries, including more negative ones than positive ones, yields better results than not expanding it at all.
Thus, we need to force the selector to pay more attention to selecting the positive ones. Therefore, we assign higher weights to positive queries than to negative ones.
This is achieved by assigning a weight to each class equal to
$w[y] = \frac{|\text{negative class}|}{|\text{class y}|}$.
Then the learning objective of the selector is as Eq.~\ref{eq: weight assignment}, in which the weight assignment regularizes the model to focus more on the positive category and less on the negative category. 
\begin{equation}
\label{eq: weight assignment}
    \mathcal{L}_{\text{S-W.A.}} = - w[y] \cdot \mathbb I\{L_{x_i}^{\text{pred.}}, L_{x_i}^{\text{gold}}\} \cdot \log\frac{\exp(x_{i},y)}{\sum_{j=1}^Y \exp(x_{i},j)}.
\end{equation}

\subsubsection{Conversational  Retrieval with Selective Expansion}

After selector training, we can apply it to select useful expansion historical queries. According to the relevant judgment produced by the selector, the current query turn can be reformulated by expanding the predicted positive queries and fed into the (off-the-shelf) retriever (ANCE) to produce a rank list. Such a pipeline takes each reformulated query turn as a stand-alone query, then any off-the-shelf retriever can be directly applied without specifically re-training for the conversational scenarios.

\subsection{Multi-Task Learning for Jointly Selector-Retriever (S-R) Fine-tuning}

We want to use conversational search data to fine-tune the retriever. We expect that the fine-tuned retriever would perform better. However, our preliminary experiment indicates that directly using the selected expansion queries as the input for the fine-tuning would degrade the final performance of the fine-tuned retriever (Sec.~\ref{sec: Conversational Finetune}).
Thus, a different strategy used for leveraging the automatically labeled queries should be designed.

From the retrieval perspective, a historical query might be always potentially relevant to the current turn to some extent, depending on the fine-tuning stage of the retriever. In the previous strategy, we only keep the queries with positive labels and remove the others. This "hard" selection strategy is static and does not allow for changes during fine-tuning. Instead, we will use a "soft" weighting of historical queries, so that during fine-tuning, the status of a historical query may change depending on the fine-tuned retriever.
In other words, we want to make the relevance labeling dependent on the retriever used. 

To implement this idea, we propose a multi-task learning method to jointly fine-tune the selector and retriever using conversational search data.
We take the query $q^{\text{all}}$ concatenating with all the historical queries as input and train a way to assign weights to each candidate query, while also trying to optimize the retrieval effectiveness.
This corresponds to a "soft" selection of candidate expansion queries embedded in the retrieval process. The expected advantage is that the retriever is optimized together with the soft selection of expansion queries, thereby alleviating the problem of data scarcity and yielding effective fine-tuning.

The joint training objective consists of two parts. One is the contrastive learning ranking loss for the retriever as shown in Eq.~\ref{eq: R}.
Another one is the loss of selector training with the weight assignment strategy. The intuition is that if the retriever has the ability to distinguish relevant queries, it would be able to assign reasonable weights to the candidate expansion queries in $q^{\text{all}}$. The whole learning object of S-R is shown as Eq.~\ref{eq: multi-task F-R}. Since retrieval is the main task, $\alpha$ aims to moderate the effect given to selector training.
\begin{align}
    \mathcal{L}_{R} &= -\log \frac{\exp\left(q_n^{\text{all}} \cdot p_n^+\right)}{\exp\left(q_n^{\text{all}} \cdot p_n^+\right) + \sum_{p_n^- \in D^-} \exp\left(q_n^{\text{all}} \cdot p_n^-\right)},
    \label{eq: R} \\ 
    \mathcal{L}_{\text{S-R}} &= \alpha \cdot \mathcal{L}_{\text{S-W.A.}} + \mathcal{L}_{R}.
    \label{eq: multi-task F-R}
\end{align}

\section{Experimental Setup}
\label{sec: experimental setup}

\subsection{Datasets and Evaluation Metrics}

\textbf{\textit{Datasets}}: We use four conversational search datasets in our experiments.
The TopiOCQA~\cite{adlakha2022topiocqa} and QReCC~\cite{anantha2021open} with relatively large amounts of training samples are used for evaluating our supervised methods (i.e. with selector training), while the other two small datasets CAsT 2019~\cite{dalton2020trec} and CAsT 2020~\cite{dalton2021cast}
are used for evaluating zero-shot retrieval (i.e. no further training). More details about the datasets are provided in Appendix~\ref{appendix: datasets}.




\noindent \textbf{\textit{Evaluation}}: Following the the existing studies on conversational search~\cite{adlakha2022topiocqa,anantha2021open,dalton2020trec,dalton2021cast}. We evaluate the retrieval effectiveness with the metrics: MRR, NDCG@3, Recall@10, Recall@20, and Recall@100. 
We adopt the \textit{pytrec\_eval} tool~\cite{sigir18_pytrec_eval} for metric computation.

\subsection{Baselines}
Since the experimental settings of existing works are not exactly the same as ours, it is unfair to directly compare them.
Thus, we implement several strong baselines following the existing methods with the conditions described in Sec.~\ref{sec: Task Formulation}.
The comparisons for supervised methods include: (1) \textbf{No expand}: The query of the current turn without expansions. 
(2) \textbf{QuReTeC}~\cite{voskarides2020query}: A token-level expansion method to train a sequence tagger to decide whether each term contained in the historical context should be added into the current query.
(3) \textbf{All expand}~\cite{yu2021few}: It expands the current query with all historical turns and input to the ANCE~\cite{xiong2020approximate} dense retriever. This is a widely used approach and the strongest baseline in the literature~\cite{qu2020open,mao2022curriculum,kim2022saving}.
(4) \textbf{Rel. expand (Human)}: It uses only the manually determined relevant historical queries to expand the current query. This method requires
human annotations about topic~\cite{adlakha2022topiocqa}, where we assume two turns are related if their ground-truth topics are the same.
(5) \textbf{Query Rewriter}~\cite{yu2020few}: This model is based on a GPT-2~\cite{radford2019language} rewriter model fine-tuned on QReCC. 

For \textit{zero-shot} setting, in addition to the \textbf{QuReTeC} and \textbf{Query Rewriter}, the other compared methods include: (6) \textbf{ZeCo}~\cite{krasakis2022zero}: A token-level representation based method using Col-BERT~\cite{khattab2020colbert} dense retriever. (7) \textbf{ANCE-ConvDR}~\cite{yu2021few}: The zero-shot version of ConvDR based on ANCE~\cite{xiong2020approximate} dense retriever. (8) \textbf{ConvDR Transfer}~\cite{yu2021few}: Fine-tune ConvDR on conversational search data only using the ranking loss (Eq.~\ref{eq: R}). 
In particular, for a fair comparison, we implement all the compared systems by training them on TopiOCQA without external datasets to make them comparable to our methods, except for the results directly quoted from original papers.

\noindent
\textbf{Oracle Methods}: We provide the retrieval evaluation by using gold PRL as the upper bound for comparison for all settings. Besides, the human-annotated relevance information from~\cite{mao2022curriculum} is provided for zero-shot scenarios for reference.

\subsection{Implementations}
We implement our selector and retriever models based on the BERT~\cite{devlin2019bert} and ANCE~\cite{xiong2020approximate} using the PyTorch and Huggingface libraries. The retriever model follows the bi-encoder architecture~\cite{karpukhin2020dense} and we fix the passage encoder during fine-tuning. More details are provided in Appendix~\ref{appendix: implementation} and our released code\footnote{\url{https://github.com/fengranMark/ConvRelExpand}}.


\section{Experiment Results}
\label{sec: experiment results}
\subsection{Dense Retrieval with Query Selector}
\label{sec: CDR with selector}

\begin{table*}[t]
    \centering
    \caption{Performance of off-the-shelf dense retrieval with the reformulated queries by the selective expansions. The human annotation is only available on TopiOCQA. The $\ddagger$ symbol denotes the significant improvements with all compared baselines in the t-test with $p < 0.05$. \textbf{Bold} and \underline{underline} indicate the best result and the second best result (other than Oracle). }
    \begin{tabular}{llcccccccc}
    \toprule
    \multirow{2}{*}{} & \multirow{2}{*}{\textbf{Method}} &
    \multicolumn{4}{c}{TopiOCQA} & \multicolumn{4}{c}{QReCC}\\
    \cmidrule(lr){3-6}\cmidrule(lr){7-10}
     & & MRR & NDCG@3 & R@20 & R@100 & MRR & NDCG@3 & R@10 & R@100\\
    \midrule
    \multirow{5}{*}{\textbf{Baseline}} & No expand & 4.08 & 3.84 & 9.67 & 13.76 & 7.50 & 6.76 & 10.77 & 17.17\\
    ~ & QuReTeC & 4.68 & 4.24 & 11.01 & 15.95 & 9.62 & 8.66 & 15.52 & 24.00 \\
    ~ & All expand & 9.09 & 8.21 & 20.64 & 29.67 & \underline{18.32} & \underline{16.78} & \underline{28.15} & \underline{40.43}\\
    ~ & Rel. expand (Human) & 10.13 & 9.20 & 22.12 & 31.58 & - & - & - & -\\
    ~ & Query Rewriter & \underline{10.72} & \underline{9.52} & \underline{23.61} & \underline{32.17} & 15.57 & 14.01 & 24.31 & 34.94\\
    \midrule
    \multirow{3}{*}{\textbf{Ours}} & Rel. expand (BERT PRL) & 9.34 & 8.74 & 19.93 & 26.89 & 16.17 & 14.80 & 25.00 & 35.03\\
    & Rel. expand (ANCE PRL) & 9.87 & 9.25 & 20.92 & 28.76 & 16.88 & 15.46 & 25.99 & 36.64\\
    & \quad + Weight Assignment & \textbf{10.83}$^\ddagger$ & \textbf{9.90}$^\ddagger$ & \textbf{24.07}$^\ddagger$ & \textbf{33.33}$^\ddagger$ & \textbf{18.53}$^\ddagger$ & \textbf{16.95}$^\ddagger$ & \textbf{28.88}$^\ddagger$ & \textbf{41.09}$^\ddagger$ \\
    \midrule
    \textbf{Oracle} & Rel. expand (Gold PRL) & 14.37$^\ddagger$ & 13.33$^\ddagger$ & 30.39$^\ddagger$ & 41.89$^\ddagger$ & 20.84$^\ddagger$ & 19.36$^\ddagger$ & 31.73$^\ddagger$ & 44.03$^\ddagger$\\
    \bottomrule
    \end{tabular}
    \label{table: PRL hard}
    \vspace{-2ex}
\end{table*}

We first evaluate the effectiveness of conversational dense retrieval with our query selection methods for query expansion. Table~\ref{table: PRL hard} shows the overall results using queries reformulated in different ways and submitted to ANCE. We show three different versions of our method: ``BERT PRL'' selects useful historical queries using BERT, ``ANCE PRL'' selects the queries using ANCE, and ``+ Weight Assignment'' uses soft selection (i.e. by assigning a weight to each historical query according to their category during training). 
We have the following observations:

(1) Our proposed methods with useful expansions outperform all baseline methods on both datasets. Specifically, our best relevant expansions (Rel. expand) improve 19.1\% and 20.6\% on MRR and NDCG@3 over the main competitor that expands with all historical context (All expand) on TopiOCQA.
The above result confirms that only part of the historical queries are useful for expanding the current query, while the remaining ones are noise that would hurt retrieval performance. Such observation is consistent with previous studies.
It also outperforms the Query Rewriter method, validating our assumption that we can reformulate the query for conversational search relying on useful historical queries automatically determined rather than manually rewritten queries. In addition, our method is much more efficient: the inference speed of our selector is hundreds of times higher than Query Rewriter. 
Besides, the method of relevant expansions by human annotations (Rel. expand Human) is also better than expanding all queries (All expand), though it is clearly lower than the oracle method and our best method with weight assignment.
This confirms our assumption that human-annotated relevant queries are not the best ones for expanding the current query, and a model can be trained to determine better useful historical queries than manual annotators. 
    
(2) The results on QReCC exhibit a similar trend to those on TopiOCQA, but the difference between the baseline and upper bound (Oracle) is smaller. Thus its performance boost is not as strong as on TopiOCQA. This is mainly because the topic-switch phenomenon is uncommon in QReCC: the conversations contain fewer turns (12.3 in TopiOCQA v.s. 3.6 in QReCC), and the queries in a conversation mostly revolve around the same topic~\cite{wu2021conqrr}. However, in a more realistic setting with a longer conversation, we expect that topic-switching in conversation would appear more often, thus the utility of our approach would be more obvious. 
    
(3) The oracle relevant expansions by gold PRL is still far ahead of that with our selector. To approach the upper bound, the PRL produced by the ANCE selector with specific strategies can further improve the retrieval scores. Thus, training a better selector is important in future work.

(4) To evaluate the performance of the selector, we report the precision, recall, macro F1, and accuracy of different selector training strategies on our generated PRL data in Table~\ref{table: PRF}. 
We can observe that there is no big difference in both the retrieval and classification performance between using BERT PRL and ANCE PRL. By exploiting the weight assignment (W.A.) strategy, all the metric scores except the recall of the selector drop a lot. Since it obtains better search results, we conjecture that the recall score of the selector is more related to the retrieval results than precision.

\subsection{Results of Conversational Fine-tuning}
\label{sec: Conversational Finetune}

\begin{table}[!t]
\centering
\small
\caption{Comparisons of Precision, Recall, Macro F1, and Accuracy of different selector training strategies.}
\setlength{\tabcolsep}{1.9mm}{
\begin{tabular}{lcccccccc}
\toprule
\multirow{2}{*}{Variants} & \multicolumn{4}{c}{TopiOCQA} & \multicolumn{4}{c}{QReCC} \\
\cmidrule(lr){2-5}\cmidrule(lr){6-9}
 & P       & \textbf{R}       &  F1 & Acc. &      P       & \textbf{R}       & F1 & Acc.    \\ \midrule
BERT PRL                & 0.55 & \textbf{0.43} & 0.48 & 0.93 & 0.51 & \textbf{0.58} & 0.54 & 0.84  \\
ANCE PRL           & 0.50 & \textbf{0.45} & 0.47 & 0.92 & 0.53 & \textbf{0.59} & 0.56 & 0.84   \\
\quad +W.A.                    & 0.26 & \textbf{0.80} & 0.39 & 0.81 & 0.37 & \textbf{0.90} & 0.52 & 0.74 \\ \midrule
Gold PRL            & 1.0     & 1.0  & 1.0     & 1.0    & 1.0    & 1.0     & 1.0     & 1.0    \\  
\bottomrule
\end{tabular}
}
\label{table: PRF}
\vspace{-2ex}
\end{table}

\begin{table}[t]
    \small
    \centering
    \caption{Performance of fine-tuned dense retrievers with conversational search data. The $\ddagger$ denotes the significant improvements with all compared baselines in the t-test with $p < 0.05$. \textbf{Bold} and \underline{underline} indicate the best result and the second best result.}
    \setlength{\tabcolsep}{1.3mm}{
    \begin{tabular}{llcccc}
    \toprule
    \multirow{2}{*}{} & \multirow{2}{*}{\textbf{Method}} &
    \multicolumn{4}{c}{TopiOCQA} \\
    \cmidrule(lr){3-6}
     & & MRR & N@3 & R@20 & R@100\\
    \midrule
    \multirow{3}{*}{\textbf{Baseline}} & No expand & 3.94 & 3.56 & 8.95 & 13.61\\
    & All expand & 9.40 & 7.90 & 24.86 & \underline{37.55}\\
    ~ & {Rel. expand (Human)} & 8.70 & 7.50 & 23.95 & 34.81\\
    \midrule
    ~ & {Rel. expand (ANCE PRL)} & 6.14 & 5.16 & 14.58 & 24.90\\
    \textbf{Ours} & {Rel. expand (Gold PRL)} & \textbf{10.26}$^\ddagger$ & \textbf{9.23}$^\ddagger$ & \textbf{26.97}$^\ddagger$ & 36.28\\
    ~ & Multi-task S-R & \underline{9.95}$^\ddagger$ & \underline{8.63}$^\ddagger$ &
    \underline{26.77}$^\ddagger$ & \textbf{39.26}$^\ddagger$\\
    \bottomrule
    \end{tabular}
    }
    \label{table: F-R Comparison}
    \vspace{-2ex}
\end{table}

In this section, we test the idea of joint fine-tuning of the selector and the retriever. 
The All expand and our Multi-task S-R method use $q^{\text{all}}$ as input for retriever fine-tuning, while the No expand and the remaining methods use $q^{\text{raw}}$ and $q^{\text{PRL}}$, respectively. Then the fine-tuned retriever might be more suitable for conversational scenarios.
The overall results are shown in Table~\ref{table: F-R Comparison}.
Together with the observations in former Sec.~\ref{sec: CDR with selector}, we have the following observations:

(1) Our multi-task S-R improves 5.5\% and 9.5\% over the strongest baseline (All expand) w.r.t MRR and NDCG@3, and approaches the performance of Rel. expand with gold PRL. This demonstrates the effectiveness of selector training as an auxiliary task. Different from the explicit selecting methods that boost MRR and NDCG@3, the implicit ones such as the All expand method and our multi-task S-R are good for Recall@100. This might be due to the fact that we use MRR as the retrieval score in Algorithm~\ref{alg: PRL}, so the Rel. expand with gold PRL will optimize MRR but not necessarily Recall@100.
    
(2) We find that the reformulated queries $q^{\text{PRL}}$ with only useful expansions cannot improve the performance of the dense retrievers after fine-tuning and even hurt it. From Table~\ref{table: F-R Comparison}, we can see the Rel. expand method with human judgment and gold PRL are worse than the All expand method on recall. The predicted PRL obtained by ANCE is even worse, indicating that the selected queries are incompatible with fine-tuning.
We also find that only All expand with $q^{\text{all}}$ and our Multi-task S-R can improve recall along with the fine-tuning procedure, while the Rel. expand method with $q^{\text{PRL}}$ keeps dropping from the beginning. Such phenomenon contradicts our intuitions and previous work~\cite{mao2022curriculum}. A possible explanation is that the gold standard PRL is produced by the base retriever before fine-tuning, while the old gold standard PRL may no longer be the gold standard for the fine-tuned retriever. 
On the other hand, our multi-task learning method can mitigate this issue to some extent. Although the gold standard PRL remains the same, the selector is updated together with the retriever and it prevents the model from performance degradation. A potentially better strategy is to update the gold PRL based on the fine-tuned retriever in every epoch so that the query relevance is estimated in tight connection with the retriever. We leave this strategy to future work.

\subsection{Zero-Shot Learning Performance}

We also evaluate our approaches in the zero-shot scenario on two CAsT datasets and compare them with other dense retrieval methods. 
The overall results are shown in Table~\ref{table: zero-shot}. 

Our methods can be divided into two parts. The selector approach is directly applying the ANCE selector obtained from Sec.~\ref{sec: CDR with selector} based on TopiOCQA, to explicitly select useful expansion turns, then feed the reformulated queries into the ANCE retriever. 
Another approach is based on fine-tuned S-R retriever obtained from Sec.~\ref{sec: Conversational Finetune} via multi-task learning. 
We can see that our methods produce higher or equivalent performance than the compared methods on various evaluation metrics. This demonstrates their high applicability to a new dataset without further fine-tuning. 
From the oracle results for reference, we can see the superior quality of both the relevant expansion by humans and our gold PRL, which indicates the high potential of the upper bound of our produced PRL data. Besides, we find the scores of human annotation are still lower than our gold PRL. This confirms again our assumption that human-annotated relevant queries are not necessarily the best ones for expansions.

\subsection{Detailed Analysis}
In this section, we make further analysis of some specific aspects of our methods.

\subsubsection{Effect of Multi-task Learning in Fine-tuning}

\begin{table}[t]
    \centering
    \small
    \caption{Performance of zero-shot dense retrieval. The $^*$ means the results
    are quoted from original papers. The $\ddagger$ denotes the significant improvements with all compared baselines in the t-test with $p < 0.05$. \textbf{Bold} and \underline{underline} indicate the best and the second best result (except Oracle).}
    \setlength{\tabcolsep}{1.8mm}{
    \begin{tabular}{lcccc}
    \toprule
    \multirow{2}{*}{\textbf{Method}} &   \multicolumn{2}{c}{CAsT-19} &   \multicolumn{2}{c}{CAsT-20}\\
    \cmidrule(lr){2-3}\cmidrule(lr){4-5}
     &  MRR &  NDCG@3 &  MRR &  NDCG@3 \\
    \midrule
    ZeCo~\cite{krasakis2022zero} & - & 23.8$^*$ & - & 17.6$^*$ \\
    ANCE-ConvDR~\cite{yu2021few} & 42.0$^*$ & 24.7$^*$ & 23.4$^*$ & 15.0$^*$ \\
    Query Rewriter~\cite{yu2020few} & 33.4 & 15.0 & 23.5 & 14.7 \\
    QuReTeC~\cite{voskarides2020query} & 45.8 & 28.2 & 21.9 & 15.3\\
    ConvDR Transfer~\cite{yu2021few} & 52.3 & 26.6 & \textbf{32.8} & \underline{20.0} \\
    \midrule
    Ours ANCE Selector & \underline{54.6}$^\ddagger$ & \textbf{31.9}$^\ddagger$ &  \textbf{32.8} & \textbf{21.7}$^\ddagger$ \\
    Ours S-R Retriever & \textbf{56.9}$^\ddagger$ & \underline{30.7}$^\ddagger$ & \underline{31.7} & 18.8 \\
    \midrule
    \multicolumn{5}{c}{Oracle - For Reference}\\
    \midrule
    \small{Rel. expand (Human)} & 65.0$^\ddagger$ & 40.8$^\ddagger$ & 42.3$^\ddagger$ & 28.8$^\ddagger$ \\
    \small{Our Rel. expand (Gold PRL)} & 70.3$^\ddagger$ & 45.1$^\ddagger$ & 47.4$^\ddagger$ & 32.4$^\ddagger$\\
    \bottomrule
    \end{tabular}
    }
    \label{table: zero-shot}
    \vspace{-2ex}
\end{table}

\begin{figure}[t]
    \centering  \includegraphics[width=\linewidth]{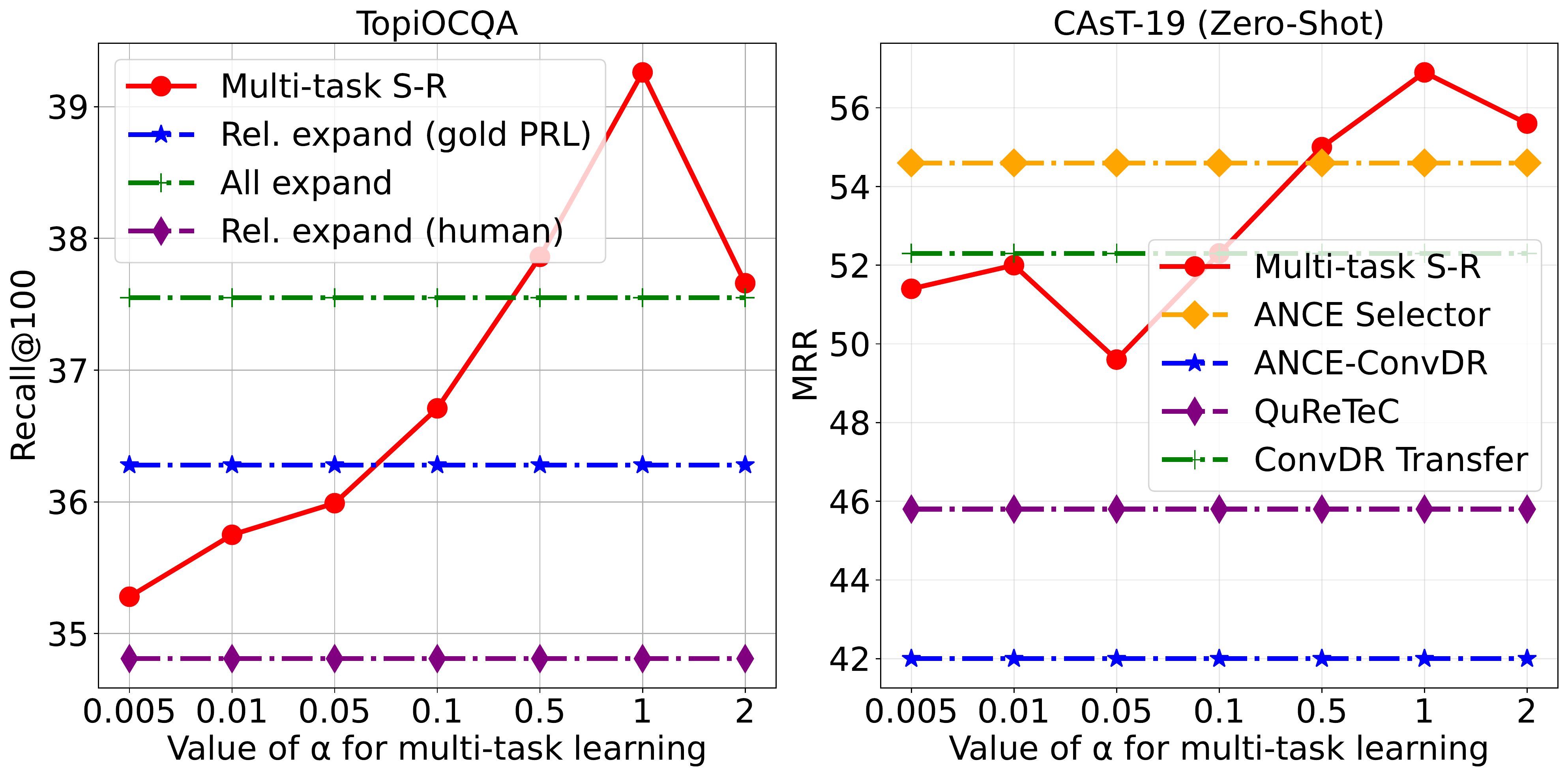}
    \caption{Analysis of S-R with multi-task learning.}
    \label{fig: Analysis alpha}
\vspace{-2ex}
\end{figure}

The selector loss $\mathcal{L}_{W.A.}$ with weight assignment plays an important role in teaching the retriever for identifying useful expansion queries during fine-tuning. We analyze the influence of hyper-parameter $\alpha$.
We first train our S-R retriever with multi-task learning on TopiOCQA then evaluate it on both TopiOCQA and CAsT-19 (zero-shot learning).
The results are shown in Fig.~\ref{fig: Analysis alpha}. We observe that the retrieval score gets better as the $\alpha$ increases until it equals one. On both datasets, $\alpha=1$ is the best setting. The results show that the two tasks in multi-task learning have about the same importance.

\begin{table}[t]
    \centering
    \small
    \caption{Analysis of topic-switch types. The  number on the left denotes the number of corresponding types. The number and percentage in parentheses denote the agreement with the human judgments on TopiOCQA. 
    }
    \vspace{-1ex}
    \scalebox{0.95}{
    \begin{tabular}{lcccccc}
    \toprule
    \multirow{2}{*}{\small{\textbf{Judgement}}} & \multicolumn{6}{c}{\textbf{Topics-Switch Type}}\\
    \cmidrule(lr){2-7}
    ~ & \multicolumn{2}{c}{\textbf{Topic Shift}} & \multicolumn{2}{c}{\textbf{Topic Return}} & \multicolumn{2}{c}{\textbf{No-switch}}\\
    \midrule
    Human & \multicolumn{2}{c}{546 (100\%)} & \multicolumn{2}{c}{126 (100\%)} & \multicolumn{2}{c}{1637 (100\%)}\\
    Pred. PRL & \multicolumn{2}{c}{501 (221, 40.48\%)} & \multicolumn{2}{c}{1313 (81, 64.29\%)} & \multicolumn{2}{c}{495 (397, 24.25\%)} \\
    Gold PRL & \multicolumn{2}{c}{1461 (469, 85.90\%)} & \multicolumn{2}{c}{581 (60, 47.62\%)} & \multicolumn{2}{c}{259 (220, 13.44\%)} \\
    \bottomrule
    \end{tabular}
    }
    \label{table: topic analysis}
    \vspace{-2ex}
\end{table}

\begin{table}[t]
    \centering
    \caption{Topic numbers by various judgment methods.}
    \vspace{-1ex}
    \begin{tabular}{lccc}
    \toprule
    \multirow{2}{*}{\textbf{Dataset}} & \multicolumn{3}{c}{\textbf{Topics / Conversation}}\\
    \cmidrule(lr){2-4}
    ~ & Human & Pred. PRL & Gold PRL\\
    \midrule
    TopiOCQA & 3.58 & 6.29 & 6.34\\
    QReCC & -  & 2.26 & 2.70\\
    CAsT-19 & 5.02 & 5.44 & 4.85\\
    CAsT-20 & 4.56 & 4.80 & 4.70\\
    \bottomrule
    \end{tabular}
    \label{table: topic number datasets}
    \vspace{-2ex}
\end{table}

\begin{figure}[t]
\centering
\includegraphics[width=\linewidth]{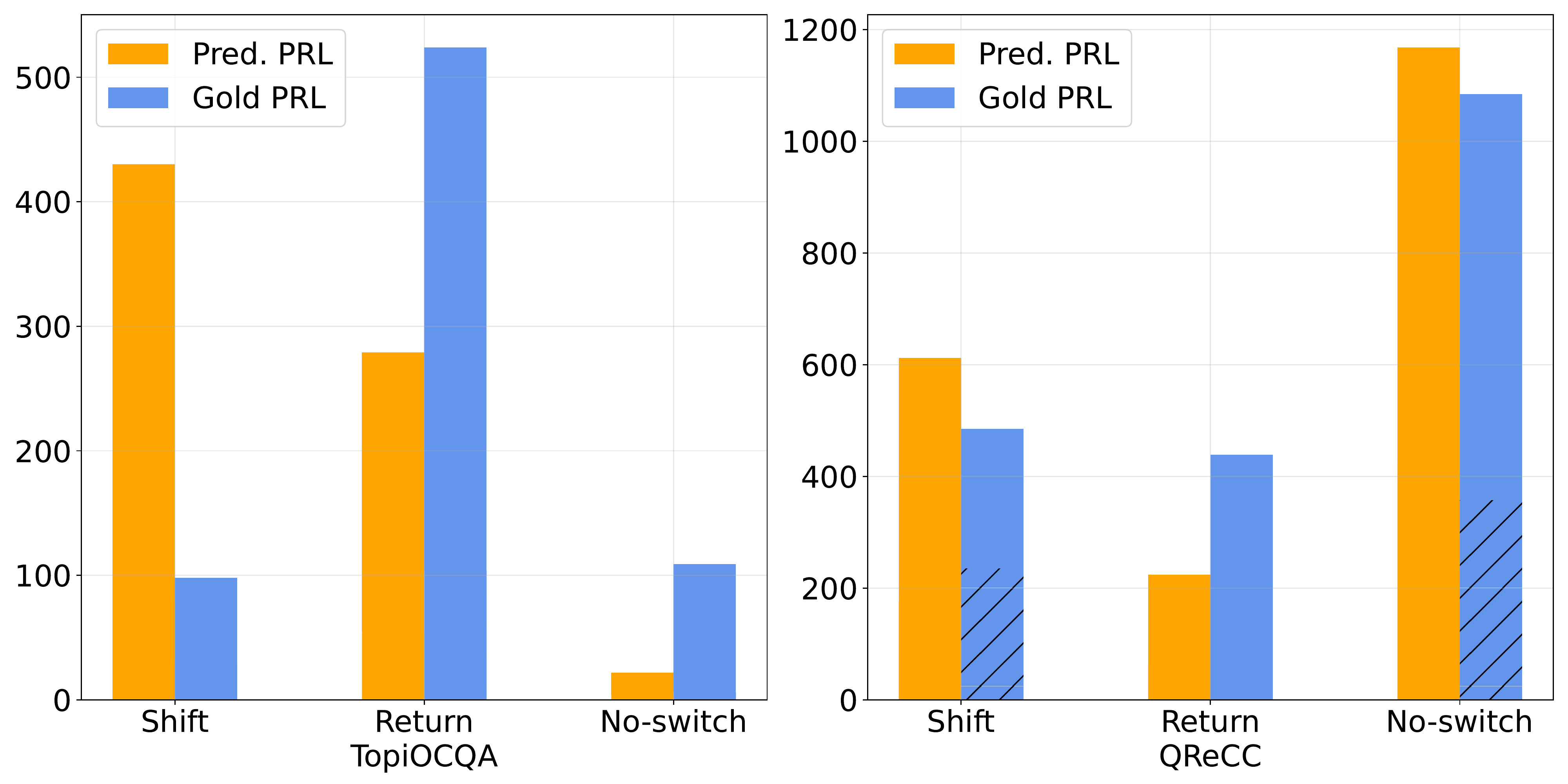}
\caption{The statistic of the successful (without shadow) and failed (with shadow) retrieval cases  under different topic-switch types and relevance judgment methods.}
\label{fig: topic}
\vspace{-2ex}
\end{figure}

\subsubsection{Effect of Topic Judgment}
Among a conversation, the turns with the same topic often have similar gold passages and thus we could assume these query turns are relevant. 
Here, we analyze the topic changes brought by different judgment methods and their impacts on retrieval performance. We follow the previous work~\cite{adlakha2022topiocqa} to classify the topic-switch into three types: topic shift, topic return, and no-switch, based on topic annotation by humans or our PRL. 
The topic shift and topic return means that the last historical query is irrelevant to the current one, while the topic return means that there are some relevant turns in history.

\begin{table*}[t]
\centering
\small
\caption{Successful and failed concrete cases about different expansion methods and their retrieval effectiveness.}
\vspace{-1ex}
\begin{tabular}{l|l}
\toprule
\multicolumn{1}{c|}{Successful Case} & \multicolumn{1}{c}{Failure Case}\\ \midrule
\begin{tabular}[c]{@{}l@{}}$\blacktriangleright$ \textbf{Context}: (TopiOCQA Session 14)\\ $q_1$: what is the man trap?\\ $r_1$: The first episode of the American science fiction television series\\ $q_2$: who are the characters? \\ $r_2$: Professor Robert Crater, his wife Nancy, Captain Kirk, Chief \\Medical Officer Dr. Leonard McCoy, and Crewman Darnell \\  $\blacktriangleright$ \textbf{Current Query}: \\ $q_3$: what is the name of the series? \\ $\blacktriangleright$ \textbf{Human Judgment}: [$q_1$, $q_2$] \quad \textbf{Pred. PRL}: [$q_1$] \quad \textbf{Gold PRL}: [$q_1$]\\
$\blacktriangleright$ \textbf{Score:} \textbf{All expand}: 0.20 \quad \textbf{Pred. PRL}: 1.0 \quad \textbf{Gold PRL}: 1.0\\
$\blacktriangleright$ \textbf{Gold Passage}: \\
$p^{*}$: ``The Man Trap'' is the first episode of the American science fiction \\ television series ... the first ``Star Trek'' episode to air on television.

\end{tabular} & \begin{tabular}[c]{@{}l@{}} 
$\blacktriangleright$ \textbf{Context}: (TopiOCQA Session 1)\\ $q_1$: when will the new dunkirk film be released on DVD? \\ $r_1$: 18 December 2017. \\ $q_2$: what is this film about? \\ $r_2$: Dunkirk evacuation of World War II\\ $\blacktriangleright$ \textbf{Current Query}: \\ $q_3$: can you mention a few members of the cast? \\ $\blacktriangleright$ \textbf{Human Judgment}: [$q_1$, $q_2$] \quad \textbf{Pred. PRL}: [] \quad \textbf{Gold PRL}: [$q_1$] \\
$\blacktriangleright$ \textbf{Score:} \textbf{All expand}: 0.33 \quad \textbf{Pred. PRL}: 0.14 \quad \textbf{Gold PRL}: 0.40\\
$\blacktriangleright$ \textbf{Gold Passage}: \\
$p^{*}$: Dunkirk is a 2017 war film written, directed, and produced \\ by Christopher Nolan that depicts the Dunkirk evacuation of \\ World War II. Its ensemble cast includes Fionn Whitehead ...
\end{tabular} \\
\bottomrule
\end{tabular}
\label{table: case study}
\vspace{-2ex}
\end{table*}

\noindent \textbf{Judgment Analysis}
Table~\ref{table: topic analysis} shows the number of topic-switch types determined in different ways:
Human, Gold PRL, and predicted PRL (Pred. PRL) by the selector. 
The difference is that the humans annotate topics with concrete topic names (e.g. "Disney Movie") for each turn, while our PRL only indicates if two queries are relevant. 
We can see that the machine judgments present more topic switches. The \textit{Gold PRL} and \textit{Pred. PRL} have considerable amounts of consistent judgments as human annotators on topic shift and topic return types. However, they do not have much overlap in no-switch judgment. As the machine-selected expansions perform better on search than that of humans, this suggests that some queries on the same topic may still be useless for expanding the current query.
This also indicates that the human topic annotation might not be the best relevant indication to use to train a selector model. 
Table~\ref{table: topic number datasets} shows the number of topics per conversation by different judgment methods. 
The number of topics in a conversation is determined by topic shift.
We can see that our PRL tends to judge more topics in TopiOCQA than in QReCC. The two CAsT datasets have slight differences between ours and human annotation. 
Previous experimental results show that the performance of our methods is much better on these three datasets except QReCC. Thus, we conjecture that our approach can make more impact on conversations with more turns, higher topic diversity, and finer-grained topics.

\noindent
\textbf{Impacts on Retrieval}
Fig.~\ref{fig: topic} shows the impact of different judgment methods on retrieval performance by counting the successful (without shadow) and failed (with shadow) examples. Here, a successful/failed example means its MRR score will increase/drop if we expand only relevant history queries to the current query turn based on the PRL, compared with taking all history queries for expansion. We find our PRL can make more impact on the topic-switch cases than those of no-switch. This phenomenon is not clearly observed on the QReCC in which the topic-switch phenomenon is less frequent.
This result is consistent with our initial intuition and previous analysis, as we might only need to expand relevant history queries if there are several topics contained in a conversation.
Another interesting finding is that the query expansion selected via Gold PRL still leads to some failure cases, which indicates the dependency between each expansion query might have an impact on retrieval effectiveness, which could be further explored. 

\subsection{Case Study}

We finally show one successful and one failed case to help understand more intuitively the behaviors of different expansion methods. We compare the results of different expansion judgments and also show the comparison of the retrieval effectiveness between expanding all historical queries and expanding only the relevant ones.

The corresponding results are shown in Table~\ref{table: case study}. For the successful case, the model is expected to understand the ``name of the series of The Man Trap''. The first round query supplies crucial information ``The Man Trap'', while the second asking for characters of the fiction would inject noise because the response for it involves names which is also the goal for the current query. Though the human annotator thinks both the first and second round queries are related to the same topic as the current one, the second round cannot provide useful information for retrieval.
For the failed case, the model needs to retrieve ``the members of the cast for the mentioned film''. Both the human judgment and the Gold PRL indicate some relevant queries in the context, and expanding them could bring improvements in retrieval effectiveness.
However, the selector did not select any expansion queries, which should be avoided and improved. The potential reason is that it is not enough to leverage only the semantics between the current query and the relevant history queries for accurate selection. Besides, the retrieval effectiveness using Gold PRL as expansion is still better than expanding all contexts. Therefore, a better selector should be able to achieve higher effectiveness for this query.

\section{Conclusion and Future Work}
In this paper, we investigated the problem of query expansion in conversational search.
We proposed a selection model to identify useful historical queries for query expansion. The model was trained using pseudo-relevance labels obtained automatically according to the impact of a candidate expansion query on retrieval effectiveness. We showed in our experiments that such a selection model can determine useful queries that perform even better than human-annotated queries. The study demonstrates that we can learn to select useful historical queries automatically without human annotation. To address the problem of inconsistency of query selection and conversational fine-tuning, we proposed a multi-task learning framework to fine-tune the selector and retriever together. This strategy produced more robust results across different datasets.

In this work, we have not explored all aspects of the method. There is much room for improvement in the future. For example, we should learn a better selector to further approach the performance of the oracle expansion. 
The differences in (topic) relevance judgments between humans and the model can be further taken into account for retrieval effectiveness. Finally, in this work, we only consider historical queries, without their retrieval results. The latter could also be used.
\begin{acks}
This work is partly supported by the National Natural Science Foundation of China (No. 61925601) and a discovery grant from the Natural Science and Engineering Research Council of Canada.
\end{acks}

\bibliographystyle{ACM-Reference-Format}
\bibliography{sample-base}


\begin{thebibliography}{44}


\ifx \showCODEN    \undefined \def \showCODEN     #1{\unskip}     \fi
\ifx \showDOI      \undefined \def \showDOI       #1{#1}\fi
\ifx \showISBNx    \undefined \def \showISBNx     #1{\unskip}     \fi
\ifx \showISBNxiii \undefined \def \showISBNxiii  #1{\unskip}     \fi
\ifx \showISSN     \undefined \def \showISSN      #1{\unskip}     \fi
\ifx \showLCCN     \undefined \def \showLCCN      #1{\unskip}     \fi
\ifx \shownote     \undefined \def \shownote      #1{#1}          \fi
\ifx \showarticletitle \undefined \def \showarticletitle #1{#1}   \fi
\ifx \showURL      \undefined \def \showURL       {\relax}        \fi
\providecommand\bibfield[2]{#2}
\providecommand\bibinfo[2]{#2}
\providecommand\natexlab[1]{#1}
\providecommand\showeprint[2][]{arXiv:#2}

\bibitem[Adlakha et~al\mbox{.}(2022)]%
        {adlakha2022topiocqa}
\bibfield{author}{\bibinfo{person}{Vaibhav Adlakha}, \bibinfo{person}{Shehzaad
  Dhuliawala}, \bibinfo{person}{Kaheer Suleman}, \bibinfo{person}{Harm de
  Vries}, {and} \bibinfo{person}{Siva Reddy}.} \bibinfo{year}{2022}\natexlab{}.
\newblock \showarticletitle{TopiOCQA: Open-domain Conversational Question
  Answering with Topic Switching}.
\newblock \bibinfo{journal}{\emph{Transactions of the Association for
  Computational Linguistics}}  \bibinfo{volume}{10} (\bibinfo{year}{2022}),
  \bibinfo{pages}{468--483}.
\newblock


\bibitem[Anantha et~al\mbox{.}(2021)]%
        {anantha2021open}
\bibfield{author}{\bibinfo{person}{Raviteja Anantha}, \bibinfo{person}{Svitlana
  Vakulenko}, \bibinfo{person}{Zhucheng Tu}, \bibinfo{person}{Shayne Longpre},
  \bibinfo{person}{Stephen Pulman}, {and} \bibinfo{person}{Srinivas Chappidi}.}
  \bibinfo{year}{2021}\natexlab{}.
\newblock \showarticletitle{Open-Domain Question Answering Goes Conversational
  via Question Rewriting}. In \bibinfo{booktitle}{\emph{Proceedings of the 2021
  Conference of the North American Chapter of the Association for Computational
  Linguistics: Human Language Technologies}}. \bibinfo{pages}{520--534}.
\newblock


\bibitem[Cao et~al\mbox{.}(2008)]%
        {cao2008selecting}
\bibfield{author}{\bibinfo{person}{Guihong Cao}, \bibinfo{person}{Jian-Yun
  Nie}, \bibinfo{person}{Jianfeng Gao}, {and} \bibinfo{person}{Stephen
  Robertson}.} \bibinfo{year}{2008}\natexlab{}.
\newblock \showarticletitle{Selecting good expansion terms for pseudo-relevance
  feedback}. In \bibinfo{booktitle}{\emph{Proceedings of the 31st annual
  international ACM SIGIR conference on Research and development in information
  retrieval}}. \bibinfo{pages}{243--250}.
\newblock


\bibitem[Dalton et~al\mbox{.}(2022)]%
        {dalton2022conversational}
\bibfield{author}{\bibinfo{person}{Jeffrey Dalton}, \bibinfo{person}{Sophie
  Fischer}, \bibinfo{person}{Paul Owoicho}, \bibinfo{person}{Filip Radlinski},
  \bibinfo{person}{Federico Rossetto}, \bibinfo{person}{Johanne~R Trippas},
  {and} \bibinfo{person}{Hamed Zamani}.} \bibinfo{year}{2022}\natexlab{}.
\newblock \showarticletitle{Conversational Information Seeking: Theory and
  Application}. In \bibinfo{booktitle}{\emph{Proceedings of the 45th
  International ACM SIGIR Conference on Research and Development in Information
  Retrieval}}. \bibinfo{pages}{3455--3458}.
\newblock


\bibitem[Dalton et~al\mbox{.}(2020)]%
        {dalton2020trec}
\bibfield{author}{\bibinfo{person}{Jeffrey Dalton}, \bibinfo{person}{Chenyan
  Xiong}, {and} \bibinfo{person}{Jamie Callan}.}
  \bibinfo{year}{2020}\natexlab{}.
\newblock \showarticletitle{TREC CAsT 2019: The conversational assistance track
  overview}.
\newblock \bibinfo{journal}{\emph{arXiv preprint arXiv:2003.13624}}
  (\bibinfo{year}{2020}).
\newblock


\bibitem[Dalton et~al\mbox{.}(2021)]%
        {dalton2021cast}
\bibfield{author}{\bibinfo{person}{Jeffrey Dalton}, \bibinfo{person}{Chenyan
  Xiong}, {and} \bibinfo{person}{Jamie Callan}.}
  \bibinfo{year}{2021}\natexlab{}.
\newblock \bibinfo{booktitle}{\emph{CAsT 2020: The Conversational Assistance
  Track Overview}}.
\newblock \bibinfo{type}{{T}echnical {R}eport}.
\newblock


\bibitem[Devlin et~al\mbox{.}(2019)]%
        {devlin2019bert}
\bibfield{author}{\bibinfo{person}{Jacob Devlin}, \bibinfo{person}{Ming-Wei
  Chang}, \bibinfo{person}{Kenton Lee}, {and} \bibinfo{person}{Kristina
  Toutanova}.} \bibinfo{year}{2019}\natexlab{}.
\newblock \showarticletitle{BERT: Pre-training of Deep Bidirectional
  Transformers for Language Understanding}. In
  \bibinfo{booktitle}{\emph{Proceedings of the 2019 Conference of the North
  American Chapter of the Association for Computational Linguistics: Human
  Language Technologies, Volume 1 (Long and Short Papers)}}.
  \bibinfo{pages}{4171--4186}.
\newblock


\bibitem[Efthimiadis(1996)]%
        {efthimiadis1996query}
\bibfield{author}{\bibinfo{person}{Efthimis~N Efthimiadis}.}
  \bibinfo{year}{1996}\natexlab{}.
\newblock \showarticletitle{Query Expansion.}
\newblock \bibinfo{journal}{\emph{Annual review of information science and
  technology (ARIST)}}  \bibinfo{volume}{31} (\bibinfo{year}{1996}),
  \bibinfo{pages}{121--87}.
\newblock


\bibitem[Gao et~al\mbox{.}(2022)]%
        {gao2022neural}
\bibfield{author}{\bibinfo{person}{Jianfeng Gao}, \bibinfo{person}{Chenyan
  Xiong}, \bibinfo{person}{Paul Bennett}, {and} \bibinfo{person}{Nick
  Craswell}.} \bibinfo{year}{2022}\natexlab{}.
\newblock \showarticletitle{Neural approaches to conversational information
  retrieval}.
\newblock \bibinfo{journal}{\emph{arXiv preprint arXiv:2201.05176}}
  (\bibinfo{year}{2022}).
\newblock


\bibitem[He and Ounis(2009)]%
        {he2009cikm}
\bibfield{author}{\bibinfo{person}{Ben He} {and} \bibinfo{person}{Iadh Ounis}.}
  \bibinfo{year}{2009}\natexlab{}.
\newblock \showarticletitle{Finding Good Feedback Documents}. In
  \bibinfo{booktitle}{\emph{Proceedings of the 18th ACM Conference on
  Information and Knowledge Management}} (Hong Kong, China)
  \emph{(\bibinfo{series}{CIKM '09})}. \bibinfo{pages}{2011–2014}.
\newblock


\bibitem[Johnson et~al\mbox{.}(2019)]%
        {johnson2019billion}
\bibfield{author}{\bibinfo{person}{Jeff Johnson}, \bibinfo{person}{Matthijs
  Douze}, {and} \bibinfo{person}{Herv{\'e} J{\'e}gou}.}
  \bibinfo{year}{2019}\natexlab{}.
\newblock \showarticletitle{Billion-scale similarity search with gpus}.
\newblock \bibinfo{journal}{\emph{IEEE Transactions on Big Data}}
  \bibinfo{volume}{7}, \bibinfo{number}{3} (\bibinfo{year}{2019}),
  \bibinfo{pages}{535--547}.
\newblock


\bibitem[Joko et~al\mbox{.}(2021)]%
        {joko2021conversational}
\bibfield{author}{\bibinfo{person}{Hideaki Joko}, \bibinfo{person}{Faegheh
  Hasibi}, \bibinfo{person}{Krisztian Balog}, {and} \bibinfo{person}{Arjen~P de
  Vries}.} \bibinfo{year}{2021}\natexlab{}.
\newblock \showarticletitle{Conversational entity linking: problem definition
  and datasets}. In \bibinfo{booktitle}{\emph{Proceedings of the 44th
  International ACM SIGIR Conference on Research and Development in Information
  Retrieval}}. \bibinfo{pages}{2390--2397}.
\newblock


\bibitem[Karpukhin et~al\mbox{.}(2020)]%
        {karpukhin2020dense}
\bibfield{author}{\bibinfo{person}{Vladimir Karpukhin}, \bibinfo{person}{Barlas
  Oguz}, \bibinfo{person}{Sewon Min}, \bibinfo{person}{Patrick Lewis},
  \bibinfo{person}{Ledell Wu}, \bibinfo{person}{Sergey Edunov},
  \bibinfo{person}{Danqi Chen}, {and} \bibinfo{person}{Wen-tau Yih}.}
  \bibinfo{year}{2020}\natexlab{}.
\newblock \showarticletitle{Dense Passage Retrieval for Open-Domain Question
  Answering}. In \bibinfo{booktitle}{\emph{Proceedings of the 2020 Conference
  on Empirical Methods in Natural Language Processing (EMNLP)}}.
  \bibinfo{pages}{6769--6781}.
\newblock


\bibitem[Khattab and Zaharia(2020)]%
        {khattab2020colbert}
\bibfield{author}{\bibinfo{person}{Omar Khattab} {and} \bibinfo{person}{Matei
  Zaharia}.} \bibinfo{year}{2020}\natexlab{}.
\newblock \showarticletitle{Colbert: Efficient and effective passage search via
  contextualized late interaction over bert}. In
  \bibinfo{booktitle}{\emph{Proceedings of the 43rd International ACM SIGIR
  conference on research and development in Information Retrieval}}.
  \bibinfo{pages}{39--48}.
\newblock


\bibitem[Kim and Kim(2022)]%
        {kim2022saving}
\bibfield{author}{\bibinfo{person}{Sungdong Kim} {and} \bibinfo{person}{Gangwoo
  Kim}.} \bibinfo{year}{2022}\natexlab{}.
\newblock \showarticletitle{Saving dense retriever from shortcut dependency in
  conversational search}. In \bibinfo{booktitle}{\emph{Proceedings of the 2022
  Conference on Empirical Methods in Natural Language Processing}}.
  \bibinfo{publisher}{Association for Computational Linguistics},
  \bibinfo{pages}{10278--10287}.
\newblock


\bibitem[Krasakis et~al\mbox{.}(2022)]%
        {krasakis2022zero}
\bibfield{author}{\bibinfo{person}{Antonios~Minas Krasakis},
  \bibinfo{person}{Andrew Yates}, {and} \bibinfo{person}{Evangelos Kanoulas}.}
  \bibinfo{year}{2022}\natexlab{}.
\newblock \showarticletitle{Zero-shot Query Contextualization for
  Conversational Search}.
\newblock \bibinfo{journal}{\emph{arXiv preprint arXiv:2204.10613}}
  (\bibinfo{year}{2022}).
\newblock


\bibitem[Kumar and Callan(2020)]%
        {2020Making}
\bibfield{author}{\bibinfo{person}{Vaibhav Kumar} {and} \bibinfo{person}{Jamie
  Callan}.} \bibinfo{year}{2020}\natexlab{}.
\newblock \showarticletitle{Making Information Seeking Easier: An Improved
  Pipeline for Conversational Search}. In \bibinfo{booktitle}{\emph{Empirical
  Methods in Natural Language Processing}}.
\newblock


\bibitem[Li et~al\mbox{.}(2022a)]%
        {li2022ditch}
\bibfield{author}{\bibinfo{person}{Huihan Li}, \bibinfo{person}{Tianyu Gao},
  \bibinfo{person}{Manan Goenka}, {and} \bibinfo{person}{Danqi Chen}.}
  \bibinfo{year}{2022}\natexlab{a}.
\newblock \showarticletitle{Ditch the Gold Standard: Re-evaluating
  Conversational Question Answering}. In \bibinfo{booktitle}{\emph{Proceedings
  of the 60th Annual Meeting of the Association for Computational Linguistics
  (Volume 1: Long Papers)}}. \bibinfo{pages}{8074--8085}.
\newblock


\bibitem[Li et~al\mbox{.}(2022b)]%
        {li2022dynamic}
\bibfield{author}{\bibinfo{person}{Yongqi Li}, \bibinfo{person}{Wenjie Li},
  {and} \bibinfo{person}{Liqiang Nie}.} \bibinfo{year}{2022}\natexlab{b}.
\newblock \showarticletitle{Dynamic Graph Reasoning for Conversational
  Open-Domain Question Answering}.
\newblock \bibinfo{journal}{\emph{ACM Transactions on Information Systems
  (TOIS)}} \bibinfo{volume}{40}, \bibinfo{number}{4} (\bibinfo{year}{2022}),
  \bibinfo{pages}{1--24}.
\newblock


\bibitem[Lin et~al\mbox{.}(2021)]%
        {lin2021contextualized}
\bibfield{author}{\bibinfo{person}{Sheng-Chieh Lin},
  \bibinfo{person}{Jheng-Hong Yang}, {and} \bibinfo{person}{Jimmy Lin}.}
  \bibinfo{year}{2021}\natexlab{}.
\newblock \showarticletitle{Contextualized Query Embeddings for Conversational
  Search}. In \bibinfo{booktitle}{\emph{Proceedings of the 2021 Conference on
  Empirical Methods in Natural Language Processing}}.
  \bibinfo{pages}{1004--1015}.
\newblock


\bibitem[Liu et~al\mbox{.}(2019)]%
        {liu2019roberta}
\bibfield{author}{\bibinfo{person}{Yinhan Liu}, \bibinfo{person}{Myle Ott},
  \bibinfo{person}{Naman Goyal}, \bibinfo{person}{Jingfei Du},
  \bibinfo{person}{Mandar Joshi}, \bibinfo{person}{Danqi Chen},
  \bibinfo{person}{Omer Levy}, \bibinfo{person}{Mike Lewis},
  \bibinfo{person}{Luke Zettlemoyer}, {and} \bibinfo{person}{Veselin
  Stoyanov}.} \bibinfo{year}{2019}\natexlab{}.
\newblock \showarticletitle{Roberta: A robustly optimized bert pretraining
  approach}.
\newblock \bibinfo{journal}{\emph{arXiv preprint arXiv:1907.11692}}
  (\bibinfo{year}{2019}).
\newblock


\bibitem[Mandya et~al\mbox{.}(2020)]%
        {mandya2020not}
\bibfield{author}{\bibinfo{person}{Angrosh Mandya}, \bibinfo{person}{James
  O’Neill}, \bibinfo{person}{Danushka Bollegala}, {and}
  \bibinfo{person}{Frans Coenen}.} \bibinfo{year}{2020}\natexlab{}.
\newblock \showarticletitle{Do not let the history haunt you: Mitigating
  Compounding Errors in Conversational Question Answering}. In
  \bibinfo{booktitle}{\emph{Proceedings of the 12th Language Resources and
  Evaluation Conference}}. \bibinfo{pages}{2017--2025}.
\newblock


\bibitem[Mao et~al\mbox{.}(2023a)]%
        {mao2023large}
\bibfield{author}{\bibinfo{person}{Kelong Mao}, \bibinfo{person}{Zhicheng Dou},
  \bibinfo{person}{Haonan Chen}, \bibinfo{person}{Fengran Mo}, {and}
  \bibinfo{person}{Hongjin Qian}.} \bibinfo{year}{2023}\natexlab{a}.
\newblock \showarticletitle{Large Language Models Know Your Contextual Search
  Intent: A Prompting Framework for Conversational Search}.
\newblock \bibinfo{journal}{\emph{arXiv preprint arXiv:2303.06573}}
  (\bibinfo{year}{2023}).
\newblock


\bibitem[Mao et~al\mbox{.}(2022a)]%
        {mao2022curriculum}
\bibfield{author}{\bibinfo{person}{Kelong Mao}, \bibinfo{person}{Zhicheng Dou},
  {and} \bibinfo{person}{Hongjin Qian}.} \bibinfo{year}{2022}\natexlab{a}.
\newblock \showarticletitle{Curriculum Contrastive Context Denoising for
  Few-shot Conversational Dense Retrieval}. In
  \bibinfo{booktitle}{\emph{Proceedings of the 45th International ACM SIGIR
  Conference on Research and Development in Information Retrieval}}.
  \bibinfo{pages}{176--186}.
\newblock


\bibitem[Mao et~al\mbox{.}(2022b)]%
        {mao2022convtrans}
\bibfield{author}{\bibinfo{person}{Kelong Mao}, \bibinfo{person}{Zhicheng Dou},
  \bibinfo{person}{Hongjin Qian}, \bibinfo{person}{Fengran Mo},
  \bibinfo{person}{Xiaohua Cheng}, {and} \bibinfo{person}{Zhao Cao}.}
  \bibinfo{year}{2022}\natexlab{b}.
\newblock \showarticletitle{ConvTrans: Transforming Web Search Sessions for
  Conversational Dense Retrieval}. In \bibinfo{booktitle}{\emph{Proceedings of
  the 2022 Conference on Empirical Methods in Natural Language Processing}}.
  \bibinfo{pages}{2935--2946}.
\newblock


\bibitem[Mao et~al\mbox{.}(2023b)]%
        {mao2023learning}
\bibfield{author}{\bibinfo{person}{Kelong Mao}, \bibinfo{person}{Hongjin Qian},
  \bibinfo{person}{Fengran Mo}, \bibinfo{person}{Zhicheng Dou},
  \bibinfo{person}{Bang Liu}, \bibinfo{person}{Xiaohua Cheng}, {and}
  \bibinfo{person}{Zhao Cao}.} \bibinfo{year}{2023}\natexlab{b}.
\newblock \showarticletitle{Learning Denoised and Interpretable Session
  Representation for Conversational Search}. In
  \bibinfo{booktitle}{\emph{Proceedings of the ACM Web Conference 2023}}.
  \bibinfo{pages}{3193--3202}.
\newblock


\bibitem[Mo et~al\mbox{.}(2023)]%
        {mo2023convgqr}
\bibfield{author}{\bibinfo{person}{Fengran Mo}, \bibinfo{person}{Kelong Mao},
  \bibinfo{person}{Yutao Zhu}, \bibinfo{person}{Yihong Wu},
  \bibinfo{person}{Kaiyu Huang}, {and} \bibinfo{person}{Jian-Yun Nie}.}
  \bibinfo{year}{2023}\natexlab{}.
\newblock \showarticletitle{ConvGQR: Generative Query Reformulation for
  Conversational Search}.
\newblock \bibinfo{journal}{\emph{arXiv preprint arXiv:2305.15645}}
  (\bibinfo{year}{2023}).
\newblock


\bibitem[Nguyen et~al\mbox{.}(2016)]%
        {nguyen2016ms}
\bibfield{author}{\bibinfo{person}{Tri Nguyen}, \bibinfo{person}{Mir
  Rosenberg}, \bibinfo{person}{Xia Song}, \bibinfo{person}{Jianfeng Gao},
  \bibinfo{person}{Saurabh Tiwary}, \bibinfo{person}{Rangan Majumder}, {and}
  \bibinfo{person}{Li Deng}.} \bibinfo{year}{2016}\natexlab{}.
\newblock \showarticletitle{MS MARCO: A human generated machine reading
  comprehension dataset}. In \bibinfo{booktitle}{\emph{CoCo@ NIPs}}.
\newblock


\bibitem[Qiao et~al\mbox{.}(2019)]%
        {qiao2019understanding}
\bibfield{author}{\bibinfo{person}{Yifan Qiao}, \bibinfo{person}{Chenyan
  Xiong}, \bibinfo{person}{Zhenghao Liu}, {and} \bibinfo{person}{Zhiyuan Liu}.}
  \bibinfo{year}{2019}\natexlab{}.
\newblock \showarticletitle{Understanding the Behaviors of BERT in Ranking}.
\newblock \bibinfo{journal}{\emph{arXiv preprint arXiv:1904.07531}}
  (\bibinfo{year}{2019}).
\newblock


\bibitem[Qu et~al\mbox{.}(2020)]%
        {qu2020open}
\bibfield{author}{\bibinfo{person}{Chen Qu}, \bibinfo{person}{Liu Yang},
  \bibinfo{person}{Cen Chen}, \bibinfo{person}{Minghui Qiu},
  \bibinfo{person}{W~Bruce Croft}, {and} \bibinfo{person}{Mohit Iyyer}.}
  \bibinfo{year}{2020}\natexlab{}.
\newblock \showarticletitle{Open-retrieval conversational question answering}.
  In \bibinfo{booktitle}{\emph{Proceedings of the 43rd International ACM SIGIR
  conference on research and development in Information Retrieval}}.
  \bibinfo{pages}{539--548}.
\newblock


\bibitem[Radford et~al\mbox{.}(2019)]%
        {radford2019language}
\bibfield{author}{\bibinfo{person}{Alec Radford}, \bibinfo{person}{Jeffrey Wu},
  \bibinfo{person}{Rewon Child}, \bibinfo{person}{David Luan},
  \bibinfo{person}{Dario Amodei}, \bibinfo{person}{Ilya Sutskever},
  {et~al\mbox{.}}} \bibinfo{year}{2019}\natexlab{}.
\newblock \showarticletitle{Language models are unsupervised multitask
  learners}.
\newblock \bibinfo{journal}{\emph{OpenAI blog}} \bibinfo{volume}{1},
  \bibinfo{number}{8} (\bibinfo{year}{2019}), \bibinfo{pages}{9}.
\newblock


\bibitem[Radlinski and Craswell(2017)]%
        {radlinski2017theoretical}
\bibfield{author}{\bibinfo{person}{Filip Radlinski} {and} \bibinfo{person}{Nick
  Craswell}.} \bibinfo{year}{2017}\natexlab{}.
\newblock \showarticletitle{A theoretical framework for conversational search}.
  In \bibinfo{booktitle}{\emph{Proceedings of the 2017 conference on conference
  human information interaction and retrieval}}. \bibinfo{pages}{117--126}.
\newblock


\bibitem[Robertson et~al\mbox{.}(2009)]%
        {robertson2009probabilistic}
\bibfield{author}{\bibinfo{person}{Stephen Robertson}, \bibinfo{person}{Hugo
  Zaragoza}, {et~al\mbox{.}}} \bibinfo{year}{2009}\natexlab{}.
\newblock \showarticletitle{The probabilistic relevance framework: BM25 and
  beyond}.
\newblock \bibinfo{journal}{\emph{Foundations and Trends{\textregistered} in
  Information Retrieval}} \bibinfo{volume}{3}, \bibinfo{number}{4}
  (\bibinfo{year}{2009}), \bibinfo{pages}{333--389}.
\newblock


\bibitem[Siblini et~al\mbox{.}(2021)]%
        {siblini2021towards}
\bibfield{author}{\bibinfo{person}{Wissam Siblini}, \bibinfo{person}{Baris
  Sayil}, {and} \bibinfo{person}{Yacine Kessaci}.}
  \bibinfo{year}{2021}\natexlab{}.
\newblock \showarticletitle{Towards a more robust evaluation for conversational
  question answering}. In \bibinfo{booktitle}{\emph{Proceedings of the 59th
  Annual Meeting of the Association for Computational Linguistics and the 11th
  International Joint Conference on Natural Language Processing (Volume 2:
  Short Papers)}}. \bibinfo{pages}{1028--1034}.
\newblock


\bibitem[Sordoni et~al\mbox{.}(2015)]%
        {sordoni2015hierarchical}
\bibfield{author}{\bibinfo{person}{Alessandro Sordoni}, \bibinfo{person}{Yoshua
  Bengio}, \bibinfo{person}{Hossein Vahabi}, \bibinfo{person}{Christina Lioma},
  \bibinfo{person}{Jakob Grue~Simonsen}, {and} \bibinfo{person}{Jian-Yun Nie}.}
  \bibinfo{year}{2015}\natexlab{}.
\newblock \showarticletitle{A hierarchical recurrent encoder-decoder for
  generative context-aware query suggestion}. In
  \bibinfo{booktitle}{\emph{proceedings of the 24th ACM international on
  conference on information and knowledge management}}.
  \bibinfo{pages}{553--562}.
\newblock


\bibitem[Vakulenko et~al\mbox{.}(2018)]%
        {vakulenko2018measuring}
\bibfield{author}{\bibinfo{person}{Svitlana Vakulenko},
  \bibinfo{person}{Maarten de Rijke}, \bibinfo{person}{Michael Cochez},
  \bibinfo{person}{Vadim Savenkov}, {and} \bibinfo{person}{Axel Polleres}.}
  \bibinfo{year}{2018}\natexlab{}.
\newblock \showarticletitle{Measuring semantic coherence of a conversation}. In
  \bibinfo{booktitle}{\emph{17th International Semantic Web Conference, ISWC
  2018}}. Springer Verlag, \bibinfo{pages}{634--651}.
\newblock


\bibitem[Vakulenko et~al\mbox{.}(2021)]%
        {vakulenko2021question}
\bibfield{author}{\bibinfo{person}{Svitlana Vakulenko}, \bibinfo{person}{Shayne
  Longpre}, \bibinfo{person}{Zhucheng Tu}, {and} \bibinfo{person}{Raviteja
  Anantha}.} \bibinfo{year}{2021}\natexlab{}.
\newblock \showarticletitle{Question rewriting for conversational question
  answering}. In \bibinfo{booktitle}{\emph{Proceedings of the 14th ACM
  International Conference on Web Search and Data Mining}}.
  \bibinfo{pages}{355--363}.
\newblock


\bibitem[Van~Gysel and de~Rijke(2018)]%
        {sigir18_pytrec_eval}
\bibfield{author}{\bibinfo{person}{Christophe Van~Gysel} {and}
  \bibinfo{person}{Maarten de Rijke}.} \bibinfo{year}{2018}\natexlab{}.
\newblock \showarticletitle{Pytrec\_eval: An Extremely Fast Python Interface to
  trec\_eval}. In \bibinfo{booktitle}{\emph{SIGIR}}. \bibinfo{publisher}{ACM}.
\newblock


\bibitem[Vaswani et~al\mbox{.}(2017)]%
        {vaswani2017attention}
\bibfield{author}{\bibinfo{person}{Ashish Vaswani}, \bibinfo{person}{Noam
  Shazeer}, \bibinfo{person}{Niki Parmar}, \bibinfo{person}{Jakob Uszkoreit},
  \bibinfo{person}{Llion Jones}, \bibinfo{person}{Aidan~N Gomez},
  \bibinfo{person}{{\L}ukasz Kaiser}, {and} \bibinfo{person}{Illia
  Polosukhin}.} \bibinfo{year}{2017}\natexlab{}.
\newblock \showarticletitle{Attention is all you need}.
\newblock \bibinfo{journal}{\emph{Advances in neural information processing
  systems}}  \bibinfo{volume}{30} (\bibinfo{year}{2017}).
\newblock


\bibitem[Voskarides et~al\mbox{.}(2020)]%
        {voskarides2020query}
\bibfield{author}{\bibinfo{person}{Nikos Voskarides}, \bibinfo{person}{Dan Li},
  \bibinfo{person}{Pengjie Ren}, \bibinfo{person}{Evangelos Kanoulas}, {and}
  \bibinfo{person}{Maarten de Rijke}.} \bibinfo{year}{2020}\natexlab{}.
\newblock \showarticletitle{Query resolution for conversational search with
  limited supervision}. In \bibinfo{booktitle}{\emph{Proceedings of the 43rd
  International ACM SIGIR conference on research and development in Information
  Retrieval}}. \bibinfo{pages}{921--930}.
\newblock


\bibitem[Wu et~al\mbox{.}(2021)]%
        {wu2021conqrr}
\bibfield{author}{\bibinfo{person}{Zeqiu Wu}, \bibinfo{person}{Yi Luan},
  \bibinfo{person}{Hannah Rashkin}, \bibinfo{person}{David Reitter}, {and}
  \bibinfo{person}{Gaurav~Singh Tomar}.} \bibinfo{year}{2021}\natexlab{}.
\newblock \showarticletitle{CONQRR: Conversational Query Rewriting for
  Retrieval with Reinforcement Learning}.
\newblock \bibinfo{journal}{\emph{arXiv preprint arXiv:2112.08558}}
  (\bibinfo{year}{2021}).
\newblock


\bibitem[Xiong et~al\mbox{.}(2020)]%
        {xiong2020approximate}
\bibfield{author}{\bibinfo{person}{Lee Xiong}, \bibinfo{person}{Chenyan Xiong},
  \bibinfo{person}{Ye Li}, \bibinfo{person}{Kwok-Fung Tang},
  \bibinfo{person}{Jialin Liu}, \bibinfo{person}{Paul Bennett},
  \bibinfo{person}{Junaid Ahmed}, {and} \bibinfo{person}{Arnold Overwijk}.}
  \bibinfo{year}{2020}\natexlab{}.
\newblock \showarticletitle{Approximate nearest neighbor negative contrastive
  learning for dense text retrieval}.
\newblock \bibinfo{journal}{\emph{arXiv preprint arXiv:2007.00808}}
  (\bibinfo{year}{2020}).
\newblock


\bibitem[Yu et~al\mbox{.}(2020)]%
        {yu2020few}
\bibfield{author}{\bibinfo{person}{Shi Yu}, \bibinfo{person}{Jiahua Liu},
  \bibinfo{person}{Jingqin Yang}, \bibinfo{person}{Chenyan Xiong},
  \bibinfo{person}{Paul Bennett}, \bibinfo{person}{Jianfeng Gao}, {and}
  \bibinfo{person}{Zhiyuan Liu}.} \bibinfo{year}{2020}\natexlab{}.
\newblock \showarticletitle{Few-shot generative conversational query
  rewriting}. In \bibinfo{booktitle}{\emph{Proceedings of the 43rd
  International ACM SIGIR conference on research and development in Information
  Retrieval}}. \bibinfo{pages}{1933--1936}.
\newblock


\bibitem[Yu et~al\mbox{.}(2021)]%
        {yu2021few}
\bibfield{author}{\bibinfo{person}{Shi Yu}, \bibinfo{person}{Zhenghao Liu},
  \bibinfo{person}{Chenyan Xiong}, \bibinfo{person}{Tao Feng}, {and}
  \bibinfo{person}{Zhiyuan Liu}.} \bibinfo{year}{2021}\natexlab{}.
\newblock \showarticletitle{Few-shot conversational dense retrieval}. In
  \bibinfo{booktitle}{\emph{Proceedings of the 44th International ACM SIGIR
  Conference on Research and Development in Information Retrieval}}.
  \bibinfo{pages}{829--838}.
\newblock


\end{thebibliography}

\appendix
\section{Appendices}
\subsection{Datasets}
\label{appendix: datasets}

The statistical information of four used conversational search datasets is shown in Table~\ref{table: Datasets}, and the details of each are described below.

\textbf{TopiOCQA} focuses on the new challenge of topic switching, which appears often in a realistic scenario.
Most conversations contain more than 10 turns and at least 3 topics. The datasets have been annotated with different types of topic switching:
topic shift - the query is a new topic, topic return - return to a previous topic, and no-switch - continue on the same topic. These annotations allow us to determine what historical queries are on the same topic as the current one. The turns involved with the same topic often have the same or similar gold passage, thus we could consider these query turns relevant. 
We consider this as the manual annotation of query relevance. This is the only dataset with such annotations.

\textbf{QReCC} focuses on the challenge of query rewriting. It aims to reformulate the query  to approach the human-rewritten query. 
Thus, it provides a gold rewritten query for each turn.
Compared with TopiOCQA, the number of turns in each conversation is smaller, and most conversations are on the same topic. We only use the queries with ground-truth for retrieval tests. 

\textbf{CAsT-19} and \textbf{CAsT-20} are two conversational search benchmarks provided in the TREC Conversational Assistance Track (CAsT). They contain 50 and 25 conversations respectively.
As we can see, no training data is provided. So we use these datasets for the setting of zero-shot evaluation, applying the models trained on another dataset (TopiOCQA).

\begin{table}[t]
    \small
    \centering
    \caption{Statistic of conversational search datasets.}
    \setlength{\tabcolsep}{1mm}{
    \begin{tabular}{lcccccc}
    \toprule
    \multirow{2}{*}{\textbf{Statistics}} & 
    \multicolumn{2}{c}{\textbf{TopiOCQA}} & 
    \multicolumn{2}{c}{\textbf{QReCC}} &
    \textbf{CAsT-19} &  \textbf{CAsT-20}\\
    \cmidrule(lr){2-3}\cmidrule(lr){4-5}\cmidrule(lr){6-6}\cmidrule(lr){7-7}
     & \textbf{Train} &  \textbf{Test} & \textbf{Train} &  \textbf{Test} & \textbf{Test} & \textbf{Test} \\
    \midrule
    \# Conversations & 3,509 & 205 & 8,987 & 2,280 & 50 & 25 \\
    \# Turns (Qry.) & 45,650 & 2,514 & 10,875 & 8,124 & 479 & 208 \\
    \# Tokens / Qry. & 6.9 & 6.9 & 6.5 & 6.4 & 6.1 & 6.8 \\
    \# Qry. / Conv. & 13.0 & 12.3 & 3.3 & 3.6 & 9.6 & 8.6 \\
    \midrule
    \#Doc in Collection & \multicolumn{2}{c}{25M} & 
    \multicolumn{2}{c}{54M} &
    \multicolumn{2}{c}{38M} \\
    \bottomrule
    \end{tabular}}
    \label{table: Datasets}
\end{table}

\subsection{Implementation Details}
\label{appendix: implementation}
Our experiments are conducted on one Nvidia A100 40G GPU. Specifically, for selector training, the length of the query pair and the batch size are both set to 128.
For retriever and multi-task S-R conversational fine-tuning, the lengths of the query, expansion concatenation, and passage are truncated into 64, 512, and 384, respectively, and the batch size is set to 16. We use Adam optimizer with 1e-5 learning rate and set the training epoch to 10. The dense retrieval is performed using Faiss~\citep{johnson2019billion}.

\end{document}